\documentclass{vldb}

\usepackage{times}
\renewcommand{\ttdefault}{cmtt}
\renewcommand{\sfdefault}{cmss}

\usepackage{booktabs} 
\usepackage{listings}
\usepackage{enumitem}\setlist{nosep}
\usepackage{subcaption}
\usepackage{adjustbox}
\usepackage{multirow}
\usepackage{siunitx}
\usepackage{balance}
\usepackage{color}
\usepackage{comment}
\usepackage{url}

\renewcommand{\footnotesize}{\small}

\renewcommand{\paragraph}[1]{\par\vspace*{0.75ex plus 0.1ex minus 0.1ex}\noindent\textbf{#1}~~}
\newcommand{\subparagraph}[1]{\par\vspace*{0.50ex plus 0.1ex minus 0.1ex}\noindent\textit{#1}~~}

\usepackage{algorithm}
\usepackage{algpseudocode}

\vldbDOI{}
\vldbTitle{}
\vldbAuthors{}
\vldbVolume{}
\vldbNumber{}
\vldbYear{}

\newtheorem{example}{Example}
\newtheorem{theorem}{Theorem}
\newtheorem{lemma}{Lemma}

\newcommand{\cut}[1]{}
\newcommand{\sql}[1]{\texttt{#1}}
\newcommand{\card}[1]{\ensuremath{\lvert#1\rvert}}
\newcommand{\Objs}{\ensuremath{\mathcal{O}}}
\newcommand{\Samples}{\ensuremath{\mathcal{S}}}
\newcommand{\SamplesI}{\ensuremath{\Samples^{\mathrm{I}}}}
\newcommand{\SamplesII}{\ensuremath{\Samples^{\mathrm{II}}}}
\newcommand{\Pred}{\ensuremath{\mathfrak{q}}}
\newcommand{\Var}{\ensuremath{\mathrm{Var}}}
\newcommand{\ceiling}[1]{\left\lceil #1 \right\rceil}

\newcommand{\Nmin}{\ensuremath{N_{\!\scalebox{0.7}{$\scriptscriptstyle\sqcup$}}}}
\newcommand{\mmin}{\ensuremath{m_{\!\scalebox{0.7}{$\scriptscriptstyle\sqcup$}}}}
\newcommand{\SamplesPS}{\ensuremath{\Gamma}}

\newcommand{\SRS}{\textsf{SRS}}
\newcommand{\SSP}{\textsf{SSP}}
\newcommand{\SSN}{\textsf{SSN}}
\newcommand{\QLCC}{\textsf{QLCC}}
\newcommand{\QLAC}{\textsf{QLAC}}
\newcommand{\QLSC}{\textsf{QLSC}}
\newcommand{\LWS}{\textsf{LWS}}
\newcommand{\LSS}{\textsf{LSS}}

\newcommand{\StraDirSol}{\textsf{DirSol}}
\newcommand{\StraLogBdr}{\textsf{LogBdr}}
\newcommand{\StraDynPgm}{\textsf{DynPgm}}
\newcommand{\PropDynPgm}{\textsf{DynPgmP}}

\newcommand{\comdots}[1]{{$\cdots$}}

\newcommand{\ansa}[1]{#1}
\newcommand{\ansb}[1]{#1}
\newcommand{\ansc}[1]{#1}
\newcommand{\ansm}[1]{#1}
\newcommand{\reva}[1]{{\ansa{#1}}}
\newcommand{\revb}[1]{{\ansb{#1}}}
\newcommand{\revc}[1]{{\ansc{#1}}}
\newcommand{\revm}[1]{{\ansm{#1}}}

\newcommand{\cutr}[1]{}   

\begin{document}

\title{Learning to Sample: Counting with Complex Queries}

\cut{
\numberofauthors{4}
\author{
\alignauthor Brett Walenz\\
	\email{bwalenz@cs.duke.edu}\\
	\affaddr{Duke University}
\alignauthor Stavros Sintos\\
	\email{ssintos@cs.duke.edu}\\
	\affaddr{Duke University}
  \and
\alignauthor Sudeepa Roy\\
	\email{sudeepa@cs.duke.edu}\\
	\affaddr{Duke University}
\alignauthor Jun Yang\\
	\email{junyang@cs.duke.edu}\\
	\affaddr{Duke University}
}
}

\numberofauthors{1}
\author{
\alignauthor Brett Walenz, Stavros Sintos, Sudeepa Roy, and Jun Yang\\
    \affaddr{Duke University, Durham, NC, USA}\\
	\email{\{bwalenz, ssintos, sudeepa, junyang\}@cs.duke.edu}
	}

\maketitle
\begin{abstract}
\begin{sloppypar}
\revm{We study the problem of efficiently estimating counts for queries involving complex filters, such as user-defined functions, or  predicates involving self-joins and correlated subqueries.  For such queries, traditional sampling techniques may not be applicable due to the complexity of the filter preventing sampling over joins, and sampling after the join may not be feasible due to
the cost of computing the full join. The other natural approach of training and using  an  inexpensive classifier to  estimate the count instead of the expensive predicate suffers from the difficulties in training a good classifier and giving meaningful confidence  intervals.  In this paper we propose a new method of \emph{learning to sample} where we combine the best of both worlds by using sampling in two phases. First, we use samples to learn a probabilistic classifier, and then use the classifier to design a stratified sampling method to  obtain the  final estimates. We theoretically analyze algorithms for obtaining an optimal stratification, and compare our approach with a suite of natural alternatives  like quantification learning, weighted  and stratified sampling, and other techniques from the literature. We also provide extensive experiments in diverse use cases using multiple real and synthetic datasets to evaluate the quality, efficiency, and robustness of our approach. }
\end{sloppypar}	
\end{abstract}

\section{Introduction}
\label{sec:intro}

Counting is a fundamental problem in query processing.  Counting
queries can be expensive to evaluate, especially if it involves
testing a complex predicate to decide whether an object should be
counted towards the total.  Consider the following example.
\begin{example}[counting points with few
  neighbors]\label{ex:k-distance}\mbox{}\newline
  Suppose table $\sql{D}(\underline{\sql{id}}, \sql{x}, \sql{y})$
  stores a set of 2d points, and we would like to count how many
  points have fewer than $k$ points within distance $d$ from them.  We
  can write the following SQL query:
  \begin{lstlisting}[basicstyle=\fontsize{7}{8}\ttfamily]
SELECT COUNT(*) FROM
(SELECT o1.id FROM D o1, D o2
 WHERE SQRT(POWER(o1.x-o2.x,2)+POWER(o1.y-o2.y,2))<=$d$
 GROUP BY o1.id HAVING COUNT(*) <= $k$);
  \end{lstlisting}
\end{example}
Here, the objects to be counted are produced by a self-join with a
complex condition, followed by \sql{GROUP BY} and \sql{HAVING}.  This
``neighborhood'' query has been well studied, with specialized index
structures and processing algorithms.  Still, there is a good chance
that a typical database system will perform poorly, either because it
has no specialized support for this query type, or it simply fails to
recognize this query type from the way the query is written.  Thus,
making such queries run faster can require a lot of effort and
expertise.  There are even more complex cases involving expensive
user-defined functions commonly found in machine learning workloads.
The problem we tackle in this paper is how to evaluate counting
queries efficiently, and in a general way.

Approximate answers are widely accepted for such expensive counting
queries.  Sampling \revb{is} a powerful technique for producing
approximate answers with statistical guarantees, with a long tradition
and active research of its applications in databases.  Yet sampling
for complex queries remains a difficult problem.  In general, not all
query operators ``commute'' with sampling.  For instance, in
Example~\ref{ex:k-distance}, if we only take a sample of \sql{D} and
evaluate the query on this sample, it would be difficult to make sense
of the result because even the neighbor counts produced by the inner
aggregation query would be off to begin with.  Worse, if the predicate
involves a black-box function with table inputs, we cannot expect
sampling input tables to produce usable results.

Still, a viable approach is to conceptually treat the problem as
counting the number of objects satisfying a predicate, where the
objects can be enumerated or sampled efficiently, but the predicate is
complex and expensive (e.g., involving user-defined functions or
arbitrarily nested subqueries).  We would sample some objects for
which we evaluate the predicate ``in full,'' and then use these
results to derive an estimate.  For instance, in
Example~\ref{ex:k-distance}, given a point \sql{o1} from \sql{D}, the
predicate would be a query over (full) \sql{D} parameterized by the
values of \sql{o1.x} and \sql{o2.x}.  Of course, evaluating the
predicate in full for each sampled object can be expensive, but
evaluating the original query as a whole can be much worse---there may
be no better way for the database systems to process this query than a
nested-loop join.  While this sampling-based approach is simple and
general, a question is whether we can make it more efficient.

Machine learning is another natural approach to this problem.  It has
the potential of being more ``sample-efficient'' because of its
ability to generalize to unseen objects.  One could draw some samples,
pay the cost to ``label'' them (i.e., evaluate the expensive
predicate), and use the labeled samples to learn a cheap classifier
that approximates the result of the expensive predicate.  The learned
classifier can then be applied to objects to obtain an estimated
count.  Beyond this naive approach, we can apply ideas from
\emph{quantification learning}~\cite{gonzalez_review_2017}.  However,
some difficulties remain: it is hard to offer meaningful statistical
guarantees (such as confidence intervals provided by sampling), and
training a good classifier can be difficult and tricky itself (e.g.,
with challenges such as feature and model selection as well as
overfitting).

A natural question is whether we can combine learning and sampling to
get the ``best of both worlds'': we want the ability to generalize by
learning, but at the same time we want the statistical guarantees
offered by sampling.  This paper answers this question in positive.
\revm{One idea is to use sampling to assess the errors produced by the
  learned classifier and correct its estimated count.  We also provide
  a novel alternative that ``learns to sample.''}  The key idea here
is not to rely directly on the learned classifier's predictions, but
instead exploit the classifier's knowledge in a more controlled manner
by using it to design a sampling scheme.  Then, we apply the sampling
scheme to derive our estimates, complete with statistical guarantees.
A good classifier leads to an efficient sampling scheme that uses few
samples to get low-variance estimates; on the other hand, a poor
classifier can lead to a less efficient sampling scheme that needs
more samples to achieve the same accuracy, but we will always have
unbiased estimates with confidence intervals.

Specifically, we make use of the scores produced by classifiers that
reflect how confident they are in their predictions.  Such scores are
readily available for popular classification methods in standard
libraries.  A straightforward method is \emph{learned weighted
  sampling}, which assigns higher sampling probabilities to objects
that are more confidently predicted to contribute to the result count.
This method is still sensitive to the scores produced by the
classifiers, and tends to focus more on confidently positive objects
instead of uncertain objects---but arguably, uncertain objects
intuitively provide more information when labeled.

Hence, we further propose \emph{learned stratified sampling}, which
relies even less on the quality of the classifier.  Instead of using
the values of the scores, we use the scores only to induce an ordering
among the objects.  Based on this ordering, and with help from some
additional samples, we find the optimal stratified sampling design
that jointly considers the partitioning of objects into strata and the
allocation of additional samples across strata.  The score-induced
ordering is useful because it brings together objects with similar
levels of uncertainty, and in particular encourages putting the
certainly positive objects and certainly negative objects into
separate strata with low within-stratum variances.  The sampling
design problem is challenging because of joint consideration of
stratification and allocation; we propose algorithms for this
optimization problem with trade-offs between speed and optimality.

Our experiments show that \revm{our learn-to-sample approach generally
  outperforms approaches that are based on either sampling or learning
  alone, or those that apply sampling only to error assessment and
  correction.}  We achieve unbiased estimates with lower variances
than other approaches, and in practice, the overhead of learning and
sampling design is negligible compared with the total cost of
evaluating expensive predicates on samples.  Moreover, learned
stratified sampling delivers robust performance even with poor
classifiers.  Finally, a key practical advantage of our
learn-to-sample approach is that it is easy to implement: its
constituent learning and sampling components are available
off-the-shelf, so we readily benefit from both the classic sampling
literature and a growing toolbox of classification algorithms.  For
example, for our experiments, we were able to apply standard
classification algorithms out-of-box with very little tuning, thanks
to the robustness of the learn-to-sample approach.


\section{Problem Definition}
\label{sec:problem}

Consider a set of objects \Objs, and a Boolean predicate
$\Pred: \Objs \to \{0,1\}$, where $1$ denotes true.  We say an object
$o$ is \emph{positive} if $\Pred(o) = 1$, or \emph{negative} if
$\Pred(o) = 0$.  Our goal is to estimate $C(\Objs, \Pred)$, the number
of positive objects in $\Objs$; i.e.,
$C(\Objs, \Pred) = \sum_{o \in \Objs} \Pred(o)$.

In general, each object $o$ can have a complex structure (with
multiple attributes including set-valued ones), and $\Pred(o)$ can be
arbitrarily complex (e.g., accessing related information beyond the
contents of $o$, comparing $o$ with other objects in \Objs, etc.).

We make two assumptions: 1)~evaluation of $\Pred$ is costly;
2)~members of \Objs\ can be efficiently enumerated.  The terms
``costly'' and ``efficient,'' of course, are relative.  While the
techniques in this paper do not depend on these assumptions for
correctness, our proposed approach is intended for situations where
these assumptions hold.  For example, a costly $\Pred$ would make it
attractive to use sampling to avoid evaluating $\Pred$ for all
objects, or to use a learned model that predicts the outcome of
$\Pred$ at a lower cost.

It should be obvious that the problem formulation above handles
single-table selection queries whose conditions potentially involve
expensive user-defined functions.  The problem formulation is also
general enough to capture more complex queries.  The first example
below illustrates the case where $\Pred$ is a complex SQL condition
involving an aggregate subquery\revm{; the second illustrates the case
  where $\Pred$ involves a black-box function.}

\begin{example}[$k$-skyband size]\label{ex:k-skyband}
  Consider a set of 2d points in table
  $\sql{D}(\underline{\sql{id}}, \sql{x}, \sql{y})$.  A point $p_1$
  \emph{dominates} another point $p_2$ if $p_1$'s \sql{x} and \sql{y}
  values are (resp.) no less than those of $p_2$ (i.e.,
  $p_1.\sql{x} \ge p_2.\sql{x} \wedge p_1.\sql{y} \ge p_2.\sql{y}$),
  and at least one of them is strictly greater (i.e.,
  $p_1.\sql{x} > p_2.\sql{x} \vee p_1.\sql{y} > p_2.\sql{y}$).  The
  so-called \emph{$k$-skyband} for the point set \sql{D}
  is the subset of points that are dominated by fewer than $k$ others.
  Given $o \in \sql{D}$, we define $\Pred(o)$ to test its membership
  in the $k$-skyband using the following SQL condition:
  \begin{lstlisting}[basicstyle=\fontsize{7}{8}\ttfamily]
(SELECT COUNT(*) FROM D
 WHERE x >= $o$.x AND y >= $o$.y AND (x>$o$.x OR y>$o$.y)) < $k$
  \end{lstlisting}
  Note that this predicate involves an aggregate subquery
  parameterized by $o$.  The number of points in the $k$-skyband is
  then the number of points satisfying $\Pred$.  \revb{Here, object
    enumeration is efficient (just scan \sql{D}), while predicate
    evaluation is costly in comparison (without specialized indexes).}

  Alternatively, we can write the whole $k$-skyband size query using a
  self-join and nested aggregation, without explicitly referring to
  $\Pred$:
  \begin{lstlisting}[basicstyle=\fontsize{7}{8}\ttfamily]
SELECT COUNT(*) FROM
(SELECT o1.id FROM D o1, D o2
 WHERE o2.x >= o1.x AND o2y >= o1.y
   AND (o2.x > o1.x OR o2.y > o1.y)
 GROUP BY o1.id HAVING COUNT(*) < $k$);
  \end{lstlisting}
\end{example}

\begin{example}[\revm{relevant document count}]\label{ex:doc-count}
  \revm{Consider a set of documents in table
    $\sql{D}(\underline{\sql{id}}, \sql{text})$.  Each document, based
    on the content of its \sql{text}, can be associated with zero or
    more labels from a predefined set of labels of interest.  For
    example, during electronic discovery for a legal proceeding,
    $\sql{D}$ can be a set of emails and documents, and one such label
    may indicate whether a document is in support of or against a
    particular action.  Let $\sql{labels}(\sql{text})$ denote a
    function that examines a document and returns the subset of labels
    that it is associated with.  We mark a document as highly relevant
    if it is associated with at least $k$ labels.  The following query
    returns the number of highly relevant documents:}
  \begin{lstlisting}[basicstyle=\fontsize{7}{8}\ttfamily]
SELECT COUNT(*) FROM D o
 WHERE len(labels(o.text)) >= k;
\end{lstlisting}
  \revm{Here, $\Pred$ is the \sql{WHERE} predicate, but it involves a
    complex black-box function \sql{labels} whose evaluation can be
    very expensive.  For example, if labels are highly specialized for
    a given proceeding, there may not exist good automated labeling
    procedures and we would have to evaluate \sql{labels} manually.
    In general, the predicate that determines whether a document is
    relevant can be even more complicated than counting how many
    labels it is associated with, but our problem formulation and
    solutions are designed to work with arbitrarily complicated
    $\Pred$.}
\end{example}

\paragraph{Handling More General SQL Queries}
An observant reader will notice the similarity between the last query
in Example~\ref{ex:k-skyband} and the one counting points with few
neighbors in Example~\ref{ex:k-distance}.  Despite the latter query's
lack of an explicit per-object predicate, it is not hard to see that
we can define $\Pred(o)$ for $o \in \sql{D}$ as the following complex
SQL condition involving an aggregate subquery (analogous to
Example~\ref{ex:k-skyband} above):
\begin{lstlisting}[basicstyle=\fontsize{7}{8}\ttfamily]
(SELECT COUNT(*) FROM D
 WHERE SQRT(POWER($o$.x-x,2)+POWER($o$.y-y,2)) <= $d$) <= $k$
\end{lstlisting}

More generally, suppose we are interested in counting the number of
results for the following SQL aggregate query:
\begin{lstlisting}[basicstyle=\fontsize{7}{8}\ttfamily]
SELECT $\mathbf{E}$ FROM $\mathbf{L}, \mathbf{R}$ -- (Q1)
WHERE $\theta_{\mathbf{L}}$ AND $\theta_{\mathbf{LR}}$
GROUP BY $\mathbf{G_L}$ HAVING $\phi$;
\end{lstlisting}
In the above, $\mathbf{G_L}$ is the list of group by columns,
$\mathbf{L}$ denotes the list of tables with columns in
$\mathbf{G_L}$, and $\mathbf{R}$ denotes the list of other tables in
the join with no group-by columns; $\theta_{\mathbf{L}}$ refers to the
part of the \sql{WHERE} condition that be evaluated over $\mathbf{L}$
alone, $\theta_{\mathbf{LR}}$ refers to the remaining part of the
\sql{WHERE} condition, and $\phi$ refers to the \texttt{HAVING}
condition; finally, $\mathbf{E}$ is the list of output expressions for
each group.  The problem of counting the number of results can be
formulated by defining the set \Objs\ of objects as:
\begin{lstlisting}[basicstyle=\fontsize{7}{8}\ttfamily]
SELECT DISTINCT $\mathbf{G_L}$ FROM $\mathbf{L}$ WHERE $\theta_{\mathbf{L}}$; -- (Q2)
\end{lstlisting}
and the predicate $\Pred(o)$ as:
\begin{lstlisting}[basicstyle=\fontsize{7}{8}\ttfamily]
EXISTS(SELECT $\mathbf{G_L}$ FROM $\mathbf{L}$, $\mathbf{R}$ -- (Q3)
       WHERE $\theta_{\mathbf{LR}}$ AND $\mathbf{G_L}$=$o$.*
       GROUP BY $\mathbf{G_L}$ HAVING $\phi$)
\end{lstlisting}
Again, the key takeaway is that our problem formulation is general
enough to support complex queries involving joins and aggregates
(besides the final counting).  \revb{Our approach works well as long
  as the set of objects is cheap to enumerate (i.e., the local
  selection $\theta_{\mathbf{L}}$ in (Q2) is easy to evaluate), while
  the per-object predicate (Q3) is relatively more expensive (which is
  usually the case because of join and aggregation).}


\section{Baseline Methods}
\label{sec:prelim}

We present a number baseline methods for estimating $C(\Objs, \Pred)$.
While these methods are not new, we note that some connections to our
problem (e.g., quantification learning and sampling-based data
cleaning) have never been made explicit or evaluated previously.

\subsection{Sampling-Based Methods}
\label{sec:prelim:sample}

\paragraph{Simple Random Sampling (\SRS)}
The problem of estimating $C(\Objs, \Pred)$ using sampling has been
studied extensively in the context of estimating
proportions~\cite{tille2011sampling}.  A straightforward method is
\emph{simple random sampling} (\emph{\SRS}).  Let
$\Samples \subseteq \Objs$ denote the set of $n$ objects drawn
randomly without replacement from the set $\Objs$ of all $N$ objects.
For each $o \in \Samples$, we evaluate $\Pred(o)$.  Then, an unbiased
estimator of $C(\Objs, \Pred)$ is $\hat{p} N$, where the estimated
proportion
$\hat{p} = \smash{1 \over n} \sum_{o \in \Samples} \Pred(o)$.  There
are a number of ways to derive a confidence interval for this
estimation.  The most popular one is the \emph{Wald interval}, which
approximates the error distribution using a normal distribution: the
$(1-\alpha)$ confidence interval for $\hat{p}$ in this case is
\begin{equation*}\textstyle
  \hat{p} \pm z_{\alpha/2}\sqrt{{\hat{p}(1-\hat{p}) \over n} \cdot {N-n
      \over N-1}}.
\end{equation*}
The usual caveats apply: if $\Pred$ is highly selective or highly
non-selective, the Wald interval is unreliable because normal
distribution approximation fails; one can use the more reliable
\emph{Wilson interval} instead.  See standard sampling
literature~\cite{tille2011sampling} for details.

\paragraph{Stratified Sampling (\SSP\ and \SSN)}
Stratified sampling is a method that works especially well when the
overall population can be divided into subpopulations (strata) where
objects are homogeneous within each stratum.  For example, if there is
a way to divide \Objs\ into two strata where one contains mostly
positive objects and the other contains mostly negative objects, we
can sample the two strata independently and use much fewer samples
overall than \SRS\ to achieve the same confidence interval.  The
problem, of course, is that we do not know the outcome of each
$\Pred(o)$ unless we first evaluate it.  A practical solution is to
choose some attributes of $o$ whose values are readily available and
likely correlated with the outcome of $\Pred(o)$; we can then stratify
\Objs\ according to these \emph{surrogates}.  In our case, a natural
choice for surrogates would be the attributes of $o$ used in computing
$\Pred(o)$; e.g., for Example~\ref{ex:k-distance}, we would choose
\sql{x} and \sql{y} and grid the 2d space into the desired number of
strata.

Suppose we are given a partitioning of \Objs\ into $H$ strata
$\Objs_1, \Objs_2,\linebreak \ldots, \Objs_H$, where
$N_h = \card{\Objs_h}$ denotes the size of each stratum $h$, and an
allocation of samples $n_1, n_2, \ldots, n_H$, where $n_h$ is the
number of samples allotted to stratum $h$.  Stratified sampling
randomly draws the allotted number of samples from each stratum;
denote these samples by $\Samples = \cup_{h=1}^H \Samples_h$, where
$n_h = \card{\Samples_h}$.  For each stratum $h$, using $\Samples_h$,
we can derive an unbiased estimator for the proportion $\hat{p}_h$ of
positive objects therein (as described for \SRS\ above).  Then, an
unbiased estimator of $C(\Objs, \Pred)$ is $\hat{p} N$, where
$\hat{p} = \sum_{h=1}^H W_h \hat{p}_h$ is the estimated overall
proportion and $W_h = N_h / N$ is the weight of stratum $h$.  The
variance in $\hat{p}$ is
\begin{equation}\textstyle
  \Var(\hat{p}) =
  \sum_{h=1}^H \frac{W_h^2 S_h^2}{n_h}
  - {1 \over N} \sum_{h=1}^H W_h S_h^2,\label{eq:stratified-var}
\end{equation}
where $S_h$ is the standard deviation of stratum $h$ (i.e., of the
multiset $\{ \Pred(o) \mid o \in \Objs_h\}$).  The $(1-\alpha)$
confidence interval for $\hat{p}$ is
  $\hat{p} \pm t_{\alpha/2}\scriptstyle\sqrt{\widehat{\Var}(\hat{p})}$,n
where $\widehat{\Var}(\hat{p})$ is an unbiased estimate of
$\Var(\hat{p})$ computed using~\eqref{eq:stratified-var} with $S_h^2$
substituted by an unbiased estimate from $\Samples_h$.  See standard
sampling literature~\cite{tille2011sampling} for details.

A simple strategy is \emph{proportional allocation}, where the number
of samples allotted to each stratum is proportional to its size, i.e.,
$n_h \propto N_h$.  We refer to stratified sampling with proportional
allocation as \SSP.
A more sophisticated alternative, \emph{Neyman allocation}, optimally
allocates samples according to $n_h \propto N_h S_h$, which minimizes
$\Var(\hat{p})$.  We refer to this alternative as \SSN.  In practice,
as we do not know $S_h$ in advance, \SSN\ proceeds in two stages:
\begin{enumerate}
\item Randomly draw a set $\SamplesI$ of samples to evaluate \Pred\
  with, and use them to derive an estimate of $S_h$ for each stratum
  $h$.  Then calculate the Neyman allocation using these
  estimates.\footnote{Standard caveats apply: given the desired total
    number of samples, we ensure that no stratum is allotted more
    samples than it contains, and that no stratum is allotted fewer
    than a prescribed minimum number of samples (even if its estimated
    standard deviation is close to $0$); we do so by rebalancing the
    allocation after meeting these constraints.}
\item Randomly draw the allotted number of samples from each stratum.
\end{enumerate}

\subsection{Learning-Based Methods}
\label{sec:prelim:learn}

Since $\Pred$ is expensive to evaluate, it is natural to consider
learning a binary classifier $f: \Objs \to \{0,1\}$ to approximate the
behavior of $\Pred$.  We can draw a random sample \Samples\ from
\Objs, evaluate $\Pred$ on them to obtain the ground truth, and then
use the results to train the classifier.  The classic classification
problem strives to classify each input object correctly, but for our
problem, we are concerned only with the \emph{number} of objects whose
ground-truth labels are $1$.  The resulting problem is an instance of
\emph{quantification learning}~\cite{gonzalez_review_2017}, whose goal
is to estimate the class distribution as opposed to individual labels.
While specialized algorithms are possible, it is appealing to adapt
classic classification algorithms for quantification learning, thereby
leveraging a rich palette of mature techniques.  In this section, we
first explore how, given a classifier $f$ that approximates \Pred, we
can use quantification learning to estimate $C(\Objs, \Pred)$.

We will not delve into specific classification algorithms here,
because they are not this paper's focus; our methods can work with any
of them.  For feature selection, we use a simple heuristic that
selects the attributes of $o$ referenced in \Pred, e.g., columns of
$\mathbf{L}$ referenced by $\theta_{\mathbf{LR}}$ in (Q1)
(Section~\ref{sec:problem}). \revm{We also note that training can be
  improved by \emph{active learning}~\cite{gonzalez_review_2017} as we discuss later.}

\paragraph{Classify-and-Count (\QLCC)}
A straightforward and natural approach is
\emph{Classify-and-Count}~\cite{gonzalez_review_2017}, which we refer
to as \emph{\QLCC}.  Suppose we randomly select
$\Samples \subseteq \Objs$ as training data and let
$C_\Samples = C(\Samples, \Pred)$ denote the count of positive objects
therein.  After learning $f$ from $\Samples$, we evaluate $f(o)$ for
each ``test object'' $o \in \Objs \setminus \Samples$.  Let
$C_{\text{obs}} = \sum_{o \in \Objs \setminus \Samples} f(o)$ denote
the ``observed count'' of $f$ over the test data.  We simply return
$C_{\text{obs}} + C_\Samples$ as the estimate for $C(\Objs, \Pred)$.
Should the classifier be accurate over the test data, this
estimate will be accurate as well.  However, it should be clear
that \QLCC\ is susceptible to classification errors and can
produce wildly skewed estimates when false positive/negative
counts are imbalanced.

\paragraph{Adjusted Count (\QLAC)}
To mitigate this problem, a recommended approach is \emph{Adjusted
  Count}~\cite{gonzalez_review_2017}, which we refer to as
\emph{\QLAC}.  The basic idea is to further adjust $C_{\text{obs}}$
using the rates of true and false positives estimated empirically from
the training data.  In more detail, we use $k$-fold cross validation
on the samples \Samples\ to compute $\smash{\widehat{\mathit{tpr}}}$
and $\smash{\widehat{\mathit{fpr}}}$, the estimated true and false
positive rates, respectively.  Then, we obtain an ``adjusted count''
$C_{\text{adj}}$ of $f$ over the test data by adjusting the observed
count $C_{\text{obs}}$ as follows\footnote{To see why, note that the
  proportion $\hat{p}$ of ``observed positive'' objects in the test
  data can be computed by
  $\hat{p} = p \cdot \mathit{tpr} + (1-p) \cdot \mathit{fpr}$, where
  $p$ denotes the actual positive proportion, and $\mathit{tpr}$ and
  $\mathit{fpr}$ are the true and false positive rates.  We can solve
  for $p$, and note that multiplying $\hat{p}$ and $p$ by the size of
  the test data yields the observed and actual counts.  Replacing
  $\mathit{tpr}$ and $\mathit{fpr}$ with their estimates then gives
  us~\eqref{eq:qlac}.}:
\begin{equation}\label{eq:qlac}
  C_{\text{adj}} = \frac%
  {C_{\text{obs}} - \widehat{\mathit{fpr}} \cdot \card{\Objs\!\setminus\!\Samples}}%
  {\widehat{\mathit{tpr}} - \widehat{\mathit{fpr}}}.
\end{equation}
Finally, we return $C_{\text{adj}} + C_\Samples$ as the estimate for
$C(\Objs, \Pred)$.

\paragraph{Active Learning}
 To improve the training of the classifier, we apply \emph{uncertainty
   sampling} from \emph{active learning}.  Given the high cost of
 labeling objects (evaluating \Pred), note that not all labeled objects
 are equally important to training; the idea is to prioritize labeling
 objects that the classifier is most ``uncertain'' about.  Many
 classifiers, besides predicting the class label, also compute a
 numeric score that indicates how ``confident'' they are in their
 predictions.  For our setting of a binary classifier, suppose the
 classifier provides a \emph{scoring function} $g: \Objs \to [0,1]$: if
 $g(o) = 1$ (or $0$), the classifier is totally confident in predicting
 $\Pred(o)$ to be $1$ (or $0$, resp.); a value strictly between $0$ and
 $1$, on the other hand, indicates uncertainty.  For some classifiers
 (e.g., random forest), one can intuitively interpret $g(o)$ as the
 probability that $\Pred(o)=1$, but in general, $g(o)$ may not have a
 probabilistic interpretation.  Regardless, the scoring function $g$
 gives us a way to select the ``most uncertain'' objects to label.  We
 assume that, compared with \Pred, $g$ is cheap to evaluate (in
 practice it is often a byproduct of classification).

 In more detail, suppose we have a labeling budget (in terms of the
 total number of objects on which to evaluate \Pred).  We first spend a
 portion of this budget to draw a set of objects $\Samples_0$ and train
 an initial classifier with scoring function $g_0$.  Next, we select
 another set of objects
 $\Samples_1 \subseteq \Objs \setminus \Samples_0$ according to $g_0$,
 focusing on objects that the initial classifier is most uncertain
 about.  The most straightforward method for selecting $\Samples_1$ is
 to evaluate $g_0$ for all objects in $\Objs \setminus \Samples_0$, and
 select the desired number of objects with the smallest value for
 $|g_0(o) - 0.5|$ (i.e., deviation from the ``toss-up'').  In practice,
 instead of considering all of $\Objs \setminus \Samples_0$, we
 randomly draw a large enough number of objects from
 $\Objs \setminus \Samples_0$ to evaluate $g_0$, and select among them
 objects with the smallest value for $|g_0(o) - 0.5|$.  We then
 evaluate \Pred\ for the set $\Samples_1$ of selected objects, and
 retrain the classifier using $\Samples_0 \cup \Samples_1$ as training
 data.  In general, we can augment the training data in this fashion
 multiple times until we exhaust the total labeling budget.

\begin{figure}
   \centering
   \includegraphics[width=\linewidth]{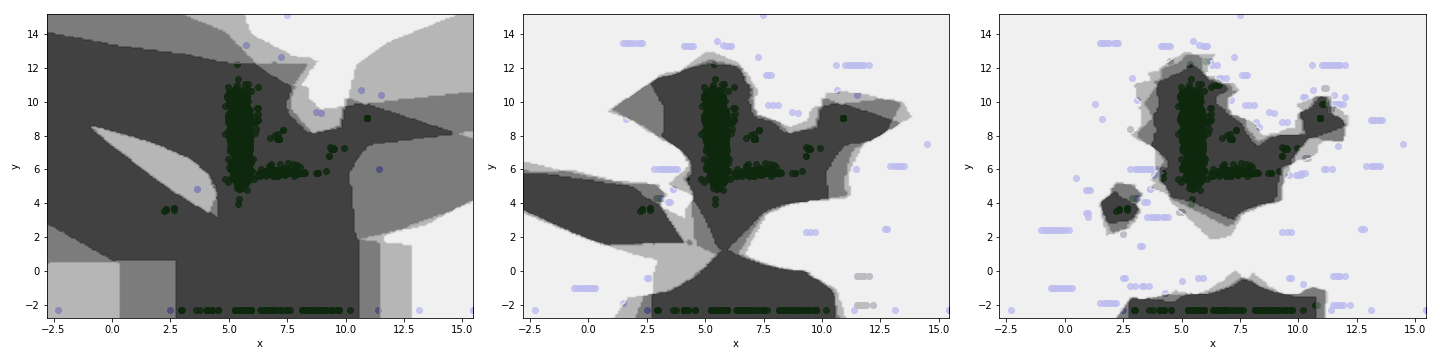}
   \caption{\label{fig:distance_augment}\small Augmenting training data
     twice for a $k$-NN classifier for Example~\ref{ex:k-distance}.
     ``$+$'' and ``$-$'' represent the training objects selected and
     their labels.  The values of scoring function $g$ over the feature
     space are represented by colors: red means the classifier
     confidently predicts $0$, blue means it confidently predicts $1$,
     while yellow means a toss-up.  From left to right, the numbers of
     training objects are $2500$ ($5\%$ of \Objs), $2600$, and $2700$.}
 \end{figure}

 As a concrete example, Figure~\ref{fig:distance_augment} shows two
 steps of augmenting the training data for Example~\ref{ex:k-distance}.
 The classifier here is a simple nearest neighbor classifier with
 \sql{x} and \sql{y} values as features.  The training data initially
 consists of $2500$ randomly drawn objects; each step adds $100$ more
 objects using the uncertainty sampling idea above.  The classifier
 scores over the feature space are shown as heat maps.  As can be seen
 intuitively from these maps, augmenting training data by drawing
 objects near the decision boundary is very effective in ``sharpening''
 the decision boundary and improving classification accuracy.

 Depending on the classification algorithm used, retraining with more
 data may add overhead, which impacts the overall efficiency (recall
 our ultimate goal of estimating $C(\Objs, \Pred)$ quickly).  As we
 have observed in our experiments, however, just one
 augmentation/retraining step gives sufficient improvement (especially
 for our new learning-to-sample methods in Section~\ref{sec:approach},
 which rely less on classifier accuracy).  Hence, we recommend a single
 augmentation/retraining step in practice, with
 $\Samples = \Samples_0 \cup \Samples_1$.

\subsection{Learning with Sample-based Correction}
\label{sec:prelim:correct}

\revb{One idea for combining learning and sampling is to follow \QLCC\
  (Classify-and-Count) with another phase, where we randomly sample
  additional objects, evaluate \Pred\ on them, assess the errors in
  the learned classifier $f$, and correct the result of
  Classify-and-Count accordingly.  We call this method
  \textbf{\itshape\QLSC}, for ``quantification learning with
  \emph{SampleClean},'' as it is inspired by the work
  of~\cite{DBLP:conf/sigmod/WangKFGKM14} on using sampling for data
  cleaning.}\footnote{While
  \emph{SampleClean}~\cite{DBLP:conf/sigmod/WangKFGKM14} deals with
  the different problem of evaluating aggregates over dirty data, its
  techniques can be adapted to our quantification learning setting by
  conceptually regarding the labels produced by the learned classifier
  as dirty data; ``cleaning'' a dirty label involves sampling the
  object and paying the cost of evaluating \Pred.  Specifically,
  \QLSC\ corresponds to their \emph{NormalizedSC} technique, which
  corrects the aggregate result computed over dirty data using the
  errors observed on data randomly selected for cleaning.  Their
  \emph{RawSC} technique, which randomly cleans data and estimates the
  result from only the cleaned labels, basically corresponds to the
  sampling-based baseline methods in our
  Section~\ref{sec:prelim:sample}.}  \revb{More precisely, recall that
  \QLCC\ samples $\Samples \subset \Objs$, learns $f$, and estimates
  the positive count over remaining objects as
  $C_{\text{obs}} = \sum_{o \in \Objs \setminus \Samples} f(o)$.
  \QLSC\ then proceeds with drawing (uniformly at random) another set
  $\Samples'$ of objects from $\Objs \setminus \Samples$, and for each
  $o \in \Samples'$ computes the error $f(o)-\Pred(o)$.  The average
  error $\hat{\epsilon}$ over $\Samples'$ gives an unbiased estimator
  for the average error over $\Objs \setminus \Samples$, so we can
  correct the count over $\Objs \setminus \Samples$ as
  $C_{\text{obs}} - \hat{\epsilon}\card{\Objs \setminus \Samples}$.
  Adding $C_\Samples$ (positive count in \Samples) yields the overall
  estimate.  Confidence intervals can be derived as in
  Section~\ref{sec:prelim:sample} because the second phase of \QLSC\
  is basically \SRS.}

\revb{\QLSC\ is similar to \QLAC\ (Section~\ref{sec:prelim:learn}) in
  that both seek to correct the result of \QLCC\ by assessing its
  errors on labeled samples.  However, \QLAC\ produces only a point
  estimate while \QLSC\ can provide confidence intervals.}


\section{Learning-to-Sample Methods}
\label{sec:approach}

In the previous section, we have seen how sampling and learning can be
applied to problem of estimating $C(\Objs, \Pred)$.  Learning is
attractive for its ability to ``generalize'' knowledge of \Pred\ to
unsampled objects, but it \revb{does not offer} the guarantees
provided by sampling (e.g., confidence intervals), and its accuracy
depends heavily on the quality of the classifier it learns.  A natural
question is whether we can combine learning and sampling to get the
``best of both worlds.''  \revm{\QLSC\
  (Section~\ref{sec:prelim:correct}) represents a baseline approach
  towards this goal: it uses sampling to correct the count predicted
  by the classifier, but its sampling scheme does not take advantage
  of the learned model in any way, and a poor classifier would result
  in a poor starting point.}

\revm{This section proposes two methods that combine learning and
  sampling more effectively.}  Both methods proceed in two phases.
\textbf{\emph{The first phase is learning}}, and is identical for the
two methods: we randomly sample objects, evaluate \Pred\ on them, and
train a binary classifier, \revm{as we did in
  Section~\ref{sec:prelim:learn}.  However, we are not going to use
  this classifier to get a count (as a starting point or otherwise).
  Instead, we assume that the classifier provides a \emph{scoring
    function} $g: \Objs \to [0,1]$: if $g(o) = 1$ (or $0$), the
  classifier is totally confident in predicting $\Pred(o)$ to be $1$
  (or $0$, resp.); a value strictly between $0$ and $1$, on the other
  hand, indicates uncertainty (e.g., $0.5$ means a toss-up).  For some
  classifiers (e.g., random forest), one can intuitively interpret
  $g(o)$ as the probability that $\Pred(o)=1$, but in general, $g(o)$
  may not have a probabilistic interpretation.  Regardless, the
  scoring function $g$ gives us a way to gauge the certainty in the
  predicted labels.  We assume that, compared with \Pred, $g$ is cheap
  to evaluate (in practice it is often a byproduct of
  classification).}

\textbf{\emph{The second phase is sampling}}, but differs between the
two methods.  The first method, \emph{Learned Weighted Sampling}
(\emph{\LWS}), is the more straightforward one of the two.  Treating
$g(o)$ has a guess of how much each object $o$ contributes to
$C(\Objs, \Pred)$, \LWS\ samples objects with higher $g(o)$ with
higher probability.  The second method, \emph{Learned Stratified
  Sampling} (\emph{LSS}), uses $g$ to guide the partitioning of
objects into strata, with the goal of reducing the variance of
estimates using stratified sampling.

The novelty of these two methods lies in their use of learning to
inform sampling.  Thanks to sampling, we still get accuracy guarantees
in the form of confidence intervals.  At the same time, we get the
benefit of learning without relying on it for correctness.  A good
classifier leads to more efficient sampling designs; on the other
hand, a poor classifier leads to a less efficient sampling design, but
we still have unbiased estimates with confidence intervals.  As we
will see, between the two methods, \LSS\ is even more robust and less
dependent on the quality of the learned classifier than \LWS.

The remainder of this section describes the second phase for these two
methods in detail.  Let $\Samples^{\text{L}}$ denote the samples used
in the first phase for learning a classifier with scoring function
$g$.  We now focus on estimating
$C(\Objs \setminus \Samples^{\text{L}}, \Pred)$ in the second phase.
In the following, we will abuse notation for simplicity: we shall
refer to $\Objs \setminus \Samples^{\text{L}}$ simply as $\Objs$
instead, and let $N = \card{\Objs}$.

\subsection{Learned Weighted Sampling}
\label{sec:approach:lws}

The second phase of \LWS\ can be seen as a form of
\emph{probability-proportional-to-size} (\emph{PPS}).  In general, PPS
relies on a ``size measure'' that is believed to be correlated to the
variable of interest.  Objects with large size measures are deemed
more important in estimation; hence, objects are drawn with
probabilities proportional to their size measures.  In our case, the
variable of interest is the result of $\Pred(o)$, so the learned
$g(o)$ can serve as the size measure.  However, to guard against an
overconfident (and potentially inaccurate) classifier, we adjust the
sampling probabilities so every $o$ has some chance of being sampled
(even if $g(o)=0$).  Specifically, we assign each $o$ an initial
sampling probability $\pi(o) \propto \max(g(o), \epsilon)$, where
$\epsilon > 0$ is a (small) prescribed threshold.  We then sample
objects from \Objs\ according to $\pi$ without replacement, evaluate
\Pred\ on the sampled objects, and estimate $C(\Objs, \Pred)$.

There are a number of estimators available from the
literature~\cite{kalton1983introduction}, including the popular
Horvitz-Thompson estimator.  We use the Des Raj estimator, whose
calculation is simpler and can provide ``ordered'' estimates, i.e.,
running estimates of mean and variance as samples are being drawn.
Let $o_1, o_2, o_3 \ldots$ denote the sequence of objects drawn
according to $\pi$ without replacement.  We compute the following
quantity after drawing each $o_i$ (with the summations below yielding
$0$ in case of $i=1$):
\begin{equation}\label{eq:resraj-i}\textstyle
  p_i = {1 \over N} \left(
    \sum_{j=1}^{i-1} \Pred(o_j) +
    {\Pred(o_i) \over \pi(o_i)} \left(1 - \sum_{j=1}^{i-1} \pi(o_j)
    \right)
  \right).
\end{equation}
The estimate for $C(\Objs, \Pred)$ after drawing the $n$-th sampled
object would be $\hat{p}^{(n)} N$, where the estimated proportion
$\hat{p}^{(n)}$ of positive objects is simply the average of all
$p_i$'s so far:
\begin{equation*}\textstyle
  \hat{p}^{(n)} = {1 \over n} \sum_{i=1}^n p_i.
\end{equation*}
And the variance in $\hat{p}^{(n)}$ can be estimated as follows:
\begin{equation*}\textstyle
  \widehat{\Var}(\hat{p}^{(n)}) = \frac{1}{n(n-1)} \sum_{i=1}^n (p_i -
  \hat{p}^{(n)})^2.
\end{equation*}
\LWS\ is very efficient when the learned classifier is accurate and
confident.  To see why, suppose the true proportion of positive
objects in \Objs\ is $p$.  For an accurate and confident classifier,
assuming an arbitrarily small $\epsilon$, $\pi(o)$ would be
arbitrarily close to $0$ if $\Pred(o) = 0$, or $\smash{1 \over pN}$
otherwise.  Therefore, each sampled object $o_i$ will have
$\Pred(o_i) = 1$ and $\pi(o_i) = \smash{1 \over pN}$.  Plugging these
into~\eqref{eq:resraj-i} and simplifying the equation yields $p_i = p$
for all $i$, so the estimate $\hat{p}^{(i)}$ at every step will be
perfectly accurate.

On the other hand, \LWS's efficiency can suffer with a poor
classifier.  Even though it still produces unbiased estimates
(regardless of the choices of $\pi(o)$'s), it may require many more
samples to achieve a tight confidence interval if it gets the
priorities wrong.

Another indication that \LWS\ may not be best for our setting is its
preference for objects with high $g(o)$.  Intuitively, focusing
instead on objects with $g(o)$ in the toss-up range reveals more
information.  Note that traditionally, PPS applies to the more general
setting where the variable of interest can be of any value; hence, it
is natural to focus on objects with potentially higher contribution to
the result.  In our setting, however, the value of interest,
$\Pred(o)$, is either $0$ or $1$.  This limited range makes our
problem easier, as we do not need to worry about cases where inclusion
or exclusion of objects with extremely high values can seriously
impact the estimates.  At the same time, this more constrained setting
also enables the possibility for better sampling designs, which we
explore next.

\subsection{Learned Stratified Sampling}
\label{sec:approach:lss}

As discussed in Section~\ref{sec:approach:lws}, the quality of the
learned classifier can adversely impact the efficiency of \LWS,
because the values of scoring function $g$ directly control the
sampling probabilities.  We now present \LSS, which uses $g$ more
conservatively, and in a way that naturally encourages exploration of
uncertain outcomes (as opposed to certain positives).

Following the learning phase, \LSS\ conceptually sorts the objects in
\Objs\ by $g$ (say, in increasing score order).  At a high level,
\LSS\ applies stratified sampling to \Objs, where stratification is
done according to this ordering; i.e., each stratum covers objects
whose $g$ scores fall into a consecutive range.  More specifically,
the second phase of \LSS\ proceeds in two stages:
\begin{enumerate}
\item Randomly draw $\SamplesI \subseteq \Objs$ to evaluate \Pred, and
  use the results to design a sampling scheme for the second
  stage---namely, the partitioning of \Objs\ into strata as well as an
  allocation of second-stage samples among these strata.
\item Randomly draw $\SamplesII \subseteq \Objs\setminus\SamplesI$ to
  evaluate \Pred, according to the sampling scheme designed by the
  first stage, and use the results to estimate $C(\Objs, \Pred)$.
\end{enumerate}
Several points are worth noting:
\begin{description}
\item[\mdseries(\emph{Versus \LWS})] While \LWS\ uses the actual $g$
  values in its sampling design, \LSS\ uses only the \emph{ordering}
  of $g$ values among objects.  Hence, \LSS\ relies less on the
  learned classifier.  We will validate this observation with
  experiments in Section~\ref{sec:expts}.  On the other hand, the
  ordering induced by $g$ is useful to \LSS\ because it intuitively
  brings together objects with similar levels of uncertainty, and in
  particular encourages putting the confidently positive objects and
  confidently negative objects into separate strata with low
  within-stratum variances.
\item[\mdseries(\emph{Versus Basic Stratified Sampling})] While the
  second phase of \LSS\ uses stratified sampling, this phase differs
  from the baseline methods in Section~\ref{sec:prelim:sample} in
  important ways: (i)~stratification in \LSS\ is based on the learned
  $g$ instead of surrogate object attributes; (ii)~\LSS\ uses
  \SamplesI\ to jointly design stratification and allocation; in
  contrast, \SSN\ only uses \SamplesI\ to design allocation (given
  stratification), while \SSP\ does not have a first stage.
\item[\mdseries(\emph{Samples in Learning and Sampling Phases})] The
  samples we draw in the sampling phase of \LSS\
  ($\SamplesI \cup \SamplesII$ above) are separate from those drawn in
  the learning phase.  Since the samples from the learning phase
  already affect (through the learned $g$) the ordering of \Objs\ for
  stratification, we choose to use new, independent samples
  (\SamplesI) for sampling design in order to minimize reliance on the
  classifier quality.\footnote{As future work, it would be interesting
    to investigate safe reuse of samples from the learning phase in
    less conservative ways.}
\end{description}

The remainder of this section discusses how we design the sampling
scheme for the second stage in detail.  Formally, we define the design
problem as follows.  Consider an ordered set \Objs\ of objects
$o_1, o_2, \ldots, o_N$ ordered by $g$ with ties broken arbitrarily, \revc{which can be  efficiently computed as we assume that the classifier is easy to execute}.
A stratification of \Objs\ into $H$ strata, specified
by $(N_1, N_2, \ldots, N_H)$ where $\sum_{h=1}^H N_h = N$, defines the
partitioning of \Objs\ into subsets $\Objs_1, \Objs_2, \ldots, \Objs_H$.
\revc{Here $\Objs_1$ includes objects with  indices $\leq N_1$, and $\Objs_h, h \geq 2$ denotes the subset of objects whose indices fall within the interval
$(\sum_{j=1}^{h-1} N_j, \sum_{j=1}^h N_j]$.
Recall from Section~\ref{sec:prelim:sample} that
\eqref{eq:stratified-var} gives the variance in the estimator of
$C(\Objs, \Pred)/N$ for stratified sampling, given the stratification $(N_1, N_2, \ldots, N_H)$ and a sample
allocation $(n_1, n_2, \ldots, n_H)$ where we draw $n_h$ objects from
$\Objs_h$.  However, we do not know the $S_h$ terms in \eqref{eq:stratified-var} in
advance, since they denote the standard deviation of the actual $\Pred(o_i)$ values of the objects $o_i \in \Objs_h$ that are expensive to compute}, so \LSS\ instead seeks to minimize the variance \revc{of $C(\Objs, \Pred)$ given by \eqref{eq:stratified-var}} estimated
using the first-stage samples \SamplesI.

More precisely, suppose the first-stage sample \SamplesI\ consists of
$m$ objects $o_{\imath_1}, o_{\imath_2}, \ldots, o_{\imath_m}$ where
$1 \le \imath_1 < \imath_2 < \cdots < \imath_m \le N$.  We aim to
find a stratification $(N_1, N_2, \ldots, N_H)$ to minimize the
objective \revc{given in \eqref{eq:stratified-obj} below  that estimates the variance in the estimator of
$C(\Objs, \Pred)$ \revc{using $n$ samples in total in the second stage.}
\revc{Here we assume $\SamplesI_h = \Objs_h \cap \SamplesI$,
                               $m_h = \card{\SamplesI_h}$, $n_h$ is number of second-stage samples in $\Objs_h$,
$\sum_{h=1}^H n_h = n$, and the variances $S_h^2$ using the first-stage samples $\SamplesI$ are estimated as
\begin{align}\textstyle
s_h^2 = {1 \over m_h-1} \sum_{o \in \SamplesI_h} (\Pred(o) -
        C(\SamplesI_h, \Pred)/m_h)^2.
        \label{eq:estimated-variance}
        \end{align}
Then the variance of the estimated $C(\Objs, \Pred)$ obtained by simplifying \eqref{eq:stratified-var}} is given by:}
\begin{align}\textstyle
  V(N_1, N_2, \ldots, N_H) &\textstyle= \sum_{h=1}^H \frac{N_h^2
                             s_h^2}{n_h} - \sum_{h=1}^H N_h
                             s_h^2.\label{eq:stratified-obj}
\end{align}
\cut{
  \text{where}\; \SamplesI_h &\textstyle= \Objs_h \cap \SamplesI,
                               \; m_h = \card{\SamplesI_h},\nonumber\\\textstyle
  s_h &\textstyle= {1 \over m_h-1} \sum_{o \in \SamplesI_h} (\Pred(o) -
        C(\SamplesI_h, \Pred)/m_h)^2,\nonumber

$n_h$ is number of second-stage samples in $\Objs_h$, and
$\sum_{h=1}^H n_h = n$.
}
The remainder of this section describes our algorithms for computing
the optimal stratification given \SamplesI. Note that the optimality
of stratification depends on the allocation strategy used.  We first
present the case of using Neyman allocation, which minimizes the
variance for a given stratification.  In this case, \LSS\ gives the
overall optimal sampling design that jointly considers stratification
and allocation.  Then, we briefly discuss the case of proportional
allocation, which is simpler but not optimal for a given
stratification.  In this case, we would find the stratification that
makes proportional allocation most effective; the optimization problem
is much easier than the case of Neyman allocation.

\subsection*{Optimizing the Stratification}

Recall from Section~\ref{sec:prelim:sample} that under Neyman
allocation using \SamplesI,
$n_h = n (N_h s_h) / (\sum_{h=1}^H N_hs_h)$.  Hence, we can further
simplify~\eqref{eq:stratified-obj}, the minimization objective, as
follows:
\begin{equation}\textstyle\label{eq:stratified-obj:neyman}
  V(\revb{N_1, N_2, \ldots, N_H}) = {1 \over n} \left( \sum_{h=1}^H N_hs_h \right)^2
  - \sum_{h=1}^H N_h s_h^2.
\end{equation}

A naive algorithm would compute $V$ for all possible stratifications \revc{$(N_1, N_2, \ldots, N_H)$}
and pick the best, but the number of possibilities is
 \revc{$\Omega(N^H)$, and computing $V$ involves
going over \SamplesI, which is expensive even for small number of partitions (e.g., when $H = 3$)}.
Before presenting our algorithms, we
describe some ideas useful to combat these challenges.

{\em First}, note that in the expression for $V$
in~\eqref{eq:stratified-obj:neyman}, \revc{from  \eqref{eq:estimated-variance}, the $s_h$ terms depend only on
the subset of objects  $\SamplesI_h$ sampled in \SamplesI\ in each stratum $h$,  and} the
precise locations of stratum boundaries between these sampled points
only affect the $N_h$ terms.  This observation suggests that we may be
able to first consider the partitioning of \SamplesI\ among strata,
and then decide where precisely the stratum boundaries lie among
\Objs.  Later in this section, we will start with an algorithm that
uses this strategy, where given the partitioning of \SamplesI, the
optimal $N_h$'s can be solved directly and (almost) exactly in the
case of $H=3$.  Building on the insights revealed in this simple case,
we then present two general algorithms for any $H$ providing
different trade-offs between speed and accuracy.  Both of these
algorithms tame complexity by restricting the potential locations of
the stratum boundaries.

{\em Second}, we can speed up the computation of $V$ significantly using
precomputation.  By sorting the $m$ objects in \SamplesI\ by $g$, we
can compute a prefix-sum index \SamplesPS, such that
$\SamplesPS(k) = \sum_{\jmath=1}^k \Pred(\smash{o_{\imath_\jmath}})$
(for $1 \le k \le m$) returns the number of positive objects among the
first $k$ objects in \SamplesI.  To obtain the indices of sampled
objects within the ordered \Objs\ (i.e.,
$\imath_1, \ldots, \imath_m$), there is no need to sort all objects in
\Objs\ by $g$.  Instead, note that the $m$ objects in \SamplesI\
divide the range of $g$ values into $m+1$ buckets; we can simply make
one pass over \Objs\ and maintain the count of objects whose $g$
values fall within each bucket.  After the pass over \Objs\ completes,
we scan the bucket counts to determine $\imath_1, \ldots, \imath_m$.

\revc{
We give the following four algorithms to compute a good stratification.
The first three algorithms work assuming the Neyman allocation, while the fourth one works for the proportional allocation.
\begin{itemize}[leftmargin=*]
\itemsep0em
  \item {\bf \StraDirSol\ (an almost optimal stratification for $H = 3$)}:
 Here  we try all pairs of \SamplesI\ as possible \emph{rough} boundaries. In particular, for each pair of consecutive samples as per $g$, we assume that the first element is the last sampled object in the first strata, while the second element is the first sampled object in the third strata. In order to find the exact boundaries in $\Objs$, we formulate and solve an optimization problem.
  \item {\bf \StraLogBdr\ (an approximate stratification for any $H$ generalizing \StraDirSol)}:
  It considers all possible ways of partitioning the $m$ sampled objects
  in \SamplesI\ among $H$ strata generalizing the ideas in \StraDirSol.
  Unlike \StraDirSol, however, for each such partitioning, we do not attempt to solve directly for the actual stratum boundaries within \Objs;
  instead, we consider only a set of candidate boundary indices, chosen judiciously to ensure that we can still find a reasonably good solution.
  In particular, between two consecutive  objects $o_{\imath_k}$ and $o_{\imath_{k+1}}$ in $\SamplesI$ (with respect to the scoring function $g$),
  we consider the objects in $\Objs$ that are $2^i$ apart from $o_{\imath_k}$ as boundary indices. 
  \item {\bf \StraDynPgm\ (a dynamic-programming-based algorithm for any $H$, which is faster than  \StraLogBdr\ but has worse approximation guarantees)}:
  A straightforward application of dynamic programming does not work since the objective function in \eqref{eq:stratified-obj:neyman} is \emph{non-separable}.
  To overcome this difficulty, we isolate the non-separable term in the objective function and solve a suite of dynamic programs where each of them operates under a different upper bound on the non-separable term.
  In order to improve the running time, we only consider as possible boundaries the set $\SamplesI$ and the additional boundary indices similar to \StraDirSol.
  In the end, we return the best result over the dynamic programs.
  \item {\bf \PropDynPgm\ (2-approximation for proportional allocation)}:
  \begin{sloppypar}
  Recall from Section~\ref{sec:prelim:sample} that under proportional allocation, $n_h = n N_h / N$. Hence, we can further
simplify \eqref{eq:stratified-obj} to $V(\revb{N_1, \ldots, N_H}) = {N-n \over n} \sum_{h=1}^H N_h s_h^2$.
The objective is much simpler than the objective for Neyman allocation and the resulting optimization problem is indeed separable, so it can be solved readily by dynamic programming. To improve the efficiency, we use the same idea as in \StraLogBdr\ and \StraDynPgm\ with  additional boundary indices.
\end{sloppypar}
\end{itemize}
}

In addition to optimizing the objective in \eqref{eq:stratified-obj:neyman}, we impose the following constraints for each stratum $h$: \revc{For two chosen thresholds $\Nmin$ and $\mmin$,}
(i) $N_h \ge \Nmin$, i.e., each stratum is large enough, and (ii) $m_h \ge \mmin$, i.e., the stratum contains enough first-stage
samples such that $s_h$ is a reasonable variance estimate.  In
practice, we have set $\mmin$ to be around $5$ and $\Nmin$ larger.

\paragraph{\StraDirSol :}
For $H=3$, we need to pick two
boundaries separating strata $\Objs_1$, $\Objs_2$, $\Objs_3$.  To this
end, suppose the last sampled object (with the largest $g$ value) in
$\Objs_1$ is the $i$-th object in \SamplesI, and the first sampled
object (with the smallest $g$ value) in $\Objs_3$ is the $j$-th object
in \SamplesI.  The algorithm considers every possible $(i, j)$ pair
where $\mmin \le i < i+\mmin < j \le m-\mmin+1$.

Given $o_{\imath_i}$ as the last sampled object in $\Objs_1$ and
$o_{\imath_j}$ as the first sampled object in $\Objs_3$, we can
readily compute $s_1, s_2, s_3$ in~\eqref{eq:stratified-obj:neyman}
using the precomputed index \SamplesPS:
$
  s_1^2 \textstyle= {\SamplesPS(i) \over i-1} \left( 1 -
          {\SamplesPS(i) \over i} \right)$,
  $s_2^2 \textstyle= {\SamplesPS(j-1)-\SamplesPS(i) \over j-i-2}
          \!\!\left(\!1\!\!-\!\!{\SamplesPS(j-1)-\SamplesPS(i) \over j-i-1}\!\!
          \right)$, and
  $s_3^2 \!\!\textstyle=\!\!{\SamplesPS(m)-\SamplesPS(j-1) \over m-j}
          \!\!\left(\!\!1\!-\!\!{\SamplesPS(m)-\SamplesPS(j-1) \over m-j+1}\!\!
          \right)$.

Then, noting that $N_2=N-N_1-N_3$, we can write $V(N_1,N_2,N_3)$ as
bivariate quadratic function $f(N_1,N_3)$ of the form
$a_1N_1^2 + a_2N_3^2 + a_3N_1N_3 + a_4N_1 + a_5N_3 + a_6$, where
coefficients $a_1,\ldots,a_6$ are computed from
$s_1,s_2,s_3,n, \text{ and } N$ (see Appendix~\ref{apndxSec1} for detailed
derivation).  Our goal is to minimize $f(N_1,N_3)$ subject to the
following constraints:
\begin{itemize}
\item $\max\{\Nmin, \imath_i\} \le N_1 \leq \imath_{i+1}-1$; i.e., the last
  sampled object in $\Objs_1$ is indeed the $i$-th one in \SamplesI,
  and $\card{\Objs_1} \ge \Nmin$.
\item $\max\{\Nmin, N-\imath_j+1\} \le N_3 \leq N-\imath_{j-1}$; i.e.,
  the first sampled object in $\Objs_3$ is the $j$-th in \SamplesI,
  and $\card{\Objs_3} \ge \Nmin$.
\item $N_1+N_3 \le N-\Nmin$; i.e., $\card{\Objs_2} \ge \Nmin$.
\end{itemize}
These constrains define a 2-dimensional polygon $R$ with at most $5$ sides.  We
optimize the function $f$ over $R$ \revc{using a standard algebraic method by considering (i) the critical
points of  $f$, and (ii) the boundary of $R$.}

To find the critical points we
set the partial derivatives of $f$ to zero and solve the resulting
linear system of two equations.  If the solution is inside $R$, we
consider it a candidate.  We then optimize $f$ for each side
($1$-facet) of $R$, which only involves optimizing a univariate
quadratic function.  We consider these solutions for the sides of $R$
as candidates too.  Finally, for each candidate, we find its closest
integer coordinate point in $R$ and evaluate $f$; we then pick the
best integer-coordinate solution.

We repeat the above procedure for each pair of sampled objects, and in
the end return the stratification with the overall minimum variance (see Appendix~\ref{apndxSec1} for additional details and the pseudocode). We call this algorithm \emph{\StraDirSol} (for direct solve).  The
following theorem summarizes its time complexity and accuracy.
\begin{theorem}\label{thm:DirSol}
  Given an ordered set \Objs\ of $N$ objects and a sampled subset
  \SamplesI\ of $m$ objects, let $v^*$ denote the minimum value of
  estimated variance defined in~\eqref{eq:stratified-obj:neyman}
  achievable using $n$ samples under stratified sampling with $H=3$
  strata where each stratum contains at least \Nmin\ objects.
  Assuming $\Nmin > n$, \StraDirSol\ runs in $O(N\log m + m^2)$ time
  and finds a stratification resulting in estimated variance
  $v \leq
  (1+\frac{2}{\Nmin}+\frac{2}{\Nmin-n}+\frac{4}{\Nmin(\Nmin-n)}) v^*$.
\end{theorem}
Note the assumption of $\Nmin>n$ above; without it, the approximation
factor would be arbitrarily bad.  In practice, however, this
assumption is weak and often holds in practice: e.g., if we take a
$5\%$ sample of \Objs\ in the second stage, this assumption means that
each stratum in \Objs\ contains at least $5\%$ of \Objs.

The algorithm is almost exact, except that the boundaries of the
strata we want to find are integers, so rounding an optimum fraction
solution to its closest integer solution may lose some accuracy.
As for running time, we can sort all objects in \SamplesI\ and precompute
\SamplesPS\ in $O(m\log m)$ time.  The indices of sampled objects
within the ordered \Objs\ can be computed in $O(N\log m)$ time, with
one pass over \Objs\ that checks each object against a balanced search
tree constructed over the $g$ values of the $m$ sampled objects.  The
algorithm considers $O(m^2)$ pairs of sampled objects, and for each
pair, it is able to minimize $f$ in $O(1)$ time, by computing
derivatives and considering only a constant number of candidate
solutions.  Therefore, overall, our algorithm takes $O(N\log m + m^2)$
time to compute the optimal stratification.

Finally, we note that this algorithm can be extended to more than $3$
strata by trying all possible size-$(H-1)$ subsets of \SamplesI\ and
optimizing an $(H-1)$-variate quadratic function subject to linear
constraints.  However, the resulting algorithm will be expensive when
$H$ and $m = \card{\SamplesI}$ are large.  In the following, we
present two less expensive approximation algorithms that work for any
$H$. \revc{The first one is slower than the second one but it has a better approximation ratio.}

\paragraph{\StraLogBdr :}
%
Given a partitioning of the sampled objects, consider
two consecutive sampled objects $o_{\imath_k}$ and $o_{\imath_{k+1}}$
that are put into different strata (there are $H-1$ such pairs of
objects).  When deciding where exactly to draw the boundary between
$o_{\imath_k}$ and $o_{\imath_{k+1}}$, the algorithm only considers
the set $B_k$ of candidate boundary indices
$\imath_k, \imath_k+2^0, \imath_k+2^1, \imath_k+2^2, \ldots$ up to
(but not including) $\imath_{k+1}$; we also add $\imath_{k+1}-1$ if it
is not already in $B_k$.  Choosing a particular index $i$ from $B_k$
means the stratum containing $o_{\imath_k}$ ends with $o_i$.  Then the
algorithm is simple.  We just check all candidate stratifications
formed by choosing one index from each of the $H-1$ sets of candidate
boundary indices.

We call this algorithm \emph{\StraLogBdr} (for logarithmic number of
candidate boundary indices).  The following theorem summarizes its
time complexity and accuracy (proof is in Appendix~\ref{apndxSec2}).
\begin{theorem}\label{thm:LogBdr}
  Given an ordered set \Objs\ of $N$ objects and a sampled subset
  \SamplesI\ of $m$ objects, let $v^*$ denote the minimum value of
  estimated variance defined in~\eqref{eq:stratified-obj:neyman}
  achievable using $n$ samples under stratified sampling with $H$
  strata where each stratum contains at least \Nmin\ objects.  Let
  $N_h^*$ denote the size of stratum $h$ in this optimum solution.
  Assuming $\Nmin > n$, \StraLogBdr\ runs in
  $O(N\log m + Hm^{H-1}\log^{H-1} N)$ time and finds a stratification
  resulting in estimated variance
  $v \leq \max\{4,\; 2+2\max_{1 \leq h \leq
    H}\smash{\frac{N_h^*}{N_h^*-n}}\} v^*$.
\end{theorem}

We can further improve the
approximation ratio if we increase the running time.
More
specifically, instead of considering candidate boundary indices in
$B_k$ that are powers of $2$ away from the sampled object index
$\imath_k$, we can consider those are powers of $(1+\epsilon)$ for a
small parameter $0 < \epsilon \leq 1$.  The approximation ratio
becomes
$\max\{{\scriptstyle (1+\epsilon)^2,\;
  (1+\epsilon)+(1+\epsilon)\max_{1 \leq h \leq H}
  \smash{\frac{N_h^*}{N_h^*-n}} }\}$ while the running time becomes
$O({\scriptstyle N\log m + \smash{\frac{H}{\epsilon}}
  m^{H-1}\log^{H-1} N})$.

\paragraph{\StraDynPgm:}
While the previous algorithm, \StraLogBdr, works for any $H$, it is
expensive due to the $m^{H-1}$ term in its running time.  While $H$ is
usually not large in practice, for a large enough $m$ (say, hundreds)
the term $m^{H-1}$ can be prohibitive even if $H=5$.  Here, we present
a faster algorithm with a larger approximation ratio that depends on
$H$.

The algorithm is based on dynamic programming.  A straightforward
application of dynamic programming would be to create an array $A$
with $N$ rows and $H$ columns, where $A[i,h]$ represents the best
we can do with $h$ strata among the first $i$ objects.  Indeed,
dynamic programming has been used previously for finding suitable
stratifications over the data, where the problems were
\emph{separable}, i.e., the solution of $A[i,h]$ could be derived by
examining the optimum solutions for $A[j,h-1]$, where $j<i$.  In our
case, however, a straightforward application would not work because
our objective function renders the problem non-separable.  To see why,
note from~\eqref{eq:stratified-obj:neyman} that
\begin{equation*}
  \begin{split}
    V(\revb{N_1, \ldots, N_H})
    &\textstyle = {1 \over n} \sum_{h=1}^H N_h^2 s_h^2 - \sum_{h=1}^H N_h s_h^2\\
    &\textstyle \; + {2 \over n} \sum_{h=2}^H \left[ N_h s_h \left(
        \sum_{h'=1}^{h-1} N_{h'} s_{h'} \right) \right].
  \end{split}
\end{equation*}
While the first two summations are separable, the third (with nesting)
is not: intuitively, the additional contribution to $V$ from the next
stratum $h$ depends on the sum $\sum_{h'=1}^{h-1} N_{h'} s_{h'}$
computed over the previous strata, but this sum is not what the
optimum solution would have minimized---we call this sum the
\emph{auxiliary sum}.

To work around the difficulty of handling the effect of this auxiliary
sum, we select a set $T$ of possible bounds on it, namely,
$T=\{2^i \mid 0 \leq i \leq \ceiling{\log(mHN)}\}$ if the auxiliary sum is greater than $1$ and $T=\{i\cdot \varepsilon \mid i\in \mathbb{Z}, i\cdot\varepsilon\leq 1\}$ for a parameter $\varepsilon$, if the sum is less than $1$. Since we do not know the value of the auxiliary sum upfront, we try all these possible values.
Then, for each
$t \in T$, we run a dynamic programming procedure operating under the
constraint that $N_hs_h \leq t$ for each $h$.  Intuitively, these
auxiliary sum constraints help us bound the quality of our solutions
even though we are not optimizing for the auxiliary sum directly.
To further reduce complexity, we also apply the same idea as in
\StraLogBdr\ to limit the set of candidate boundary indices to
consider.
Here, we will consider more indices but without increasing
the asymptotic complexity.  Specifically, for each sampled object
$o_{\imath_k} \in \SamplesI$, we consider indices
$\imath_k, \imath_k+2^0, \imath_k+2^1, \imath_k+2^2, \ldots$ up to
(but not including) $\imath_{k+1}$, as well as indices
$\imath_k-2^0, \imath_k-2^1, \imath_k-2^2, \ldots$ down to (but not
including) $\imath_{k-1}$.  We denote the ordered set of all candidate
boundary indices (induced by all sampled points) by
$B = \{ b_1, b_2, \ldots \}$.  Clearly $\card{B} = O(m \log N)$.
Furthermore, for each $b_i$ we denote by $\ell_i$ the value $k$ such
that the $k$-th sampled object is the last sampled one among
$o_1, o_2, \ldots, o_{b_i}$; we can easily record all $\ell_i$'s when
constructing $B$.

Now we can describe the dynamic programming procedure that runs for
each $t \in T$.  Let $A_t$ be an array with $\card{B}$ rows and $H$
columns, where $A_t[i,h]$ stores the variance of the best
stratification we found for $h$ strata over the first $b_i$ objects in
\Objs.  Let $X_t$ be an array of the same dimension as $A_t$, where
$X_t[i,h]$ stores the overall auxiliary sum corresponding to the
solution represented by $A_t[i,h]$.  We then have
\begin{align*}
  \begin{split}
    A_t[i,h] &= \min_{1 \leq j < i,\; N_{j,i} s_{j,i} \leq t}\\
    &\phantom{{}={}}\textstyle \big\{ A_t[j,h-1]
    + {1 \over n} N_{j,i}^2 s_{j,i}^2
    - N_{j,i} s_{j,i}^2\\
    &\phantom{{}={}}\textstyle\;\; + {2 \over n} N_{j,i} s_{j,i} X_t[j,h-1] \big\},
  \end{split}
\end{align*}
where $N_{j,i} = b_i - b_j$ is the size of stratum $h$ (containing
objects $o_{b_j+1}, \ldots, o_{b_i}$) being considered, and
$s_{j,i}^2$ is the estimated variance for stratum $h$, which can be
computed using the prefix-sum index as
$\smash{\SamplesPS(\ell_i)-\SamplesPS(\ell_j) \over \ell_i-\ell_j}
\smash{\left(1 - {\SamplesPS(\ell_i)-\SamplesPS(\ell_j) \over
      \ell_i-\ell_j+1}\right)}$.  The array entry $X_t[i,h]$ can be
updated accordingly.

After running the dynamic programming procedure for all $t \in T$, we
return the best solution found ($\min_{t\in T}A_t[\card{B},H]$).
Overall, we try $O(\log(mHN)+\frac{1}{\varepsilon})=O(\log N+\frac{1}{\varepsilon})$ values of $t$, and for each
$t$, the dynamic programming procedure takes $O(Hm^2\log^2N)$ time.
The total running time, including precomputation, is
$O(N\log m + (\log N + \frac{1}{\varepsilon})Hm^2\log^2N)$.

The proof of the next theorem can be found in Appendix~\ref{apndxSec3}.
\begin{theorem}\label{thm:DynPgm}
  Given an ordered set \Objs\ of $N$ objects and a sampled subset
  \SamplesI\ of $m$ objects, let $v^*$ denote the minimum value of
  estimated variance defined in~\eqref{eq:stratified-obj:neyman}
  achievable using $n$ samples under stratified sampling with $H$
  strata where each stratum contains at least \Nmin\ objects, and let $\varepsilon$ be any parameter with $0<\varepsilon<1$.
  Assuming $\Nmin \ge 4n$, \StraDynPgm\ runs in
  $O(N\log m + (\log N + \frac{1}{\varepsilon})Hm^2\log^2N)$ time and finds a stratification resulting
  in estimated variance $v \leq \smash{\frac{14}{3}}(10H-9) v^*$ or $v \leq \smash{\frac{14}{3}}(5H-4) v^*+\varepsilon$.
\end{theorem}
\newcommand{\remove}[1]{}
\remove{
Similarly to
\StraLogBdr, we can improve the approximation ratio by increasing the
running time.
More specifically, we can consider candidate boundary
indices that are powers of $(1+\varepsilon)$ (instead of $2$) away from
the indices of sampled objects.  We also consider more fine-grained
bounds on the auxiliary sum, namely,
$T = \{ (1+\epsilon)^i \mid 0 \leq i \leq
\ceiling{\log_{1+\epsilon}(mHN)} \} \cup \{i\cdot \varepsilon \mid i\in \mathbb{Z}, i\cdot\varepsilon\leq 1\}$.  These changes would lead to an
algorithm with an approximation ratio of
$\frac{7(1+\epsilon)}{3}\left[5(1+\epsilon)(H-1)+1\right]$ or $\smash{\frac{7(1+\varepsilon)}{3}}(6H-5) v^*+\varepsilon$ and
$O(N\log m + (\log N +\frac{1}{\varepsilon})\smash{H \over \epsilon^3} m^2\log^2 N)$ running time.
}

\paragraph{\PropDynPgm}:
\label{sec:approach:lss:stratify-prop}

Recall from Section~\ref{sec:prelim:sample} that under proportional
allocation, $n_h = n N_h / N$.  Hence, we can further
simplify~\eqref{eq:stratified-obj}, the minimization objective, as
follows:
\begin{equation}\textstyle\label{eq:stratified-obj:prop}
  V(\revb{N_1, \ldots, N_H}) = {N-n \over n} \sum_{h=1}^H N_h s_h^2.
\end{equation}
This objective is much simpler than~\eqref{eq:stratified-obj:neyman}
for Neyman allocation.  The resulting optimization problem is indeed
separable, and can be solved readily by dynamic programming.

We still carry out the same precomputation (e.g., the prefix-sum index
\SamplesPS), and we also
use the same idea behind \StraDynPgm, \StraLogBdr\ to select the ordered set of
candidate boundary indices $B = \{ b_1, b_2, \ldots \}$, where
$\card{B} = O(m \log N)$.  The dynamic programming algorithm then
proceeds as follows.  Let $A$ be an array with $\card{B}$ rows and $H$
columns, where $A[i,h]$ represents the best we can do with $h$ strata
over the first $b_i$ objects in \Objs.  We have
$A[i,h]=\min_{1 \leq j < i} \{ A[j,h-1] + {N-n \over n} N_{j,i}
s_{j,i}^2 \}$, where $N_{j,i} = b_i - b_j$ is the size of stratum $h$
being considered, and
$s_{j,i}^2 = \smash{\SamplesPS(\ell_i)-\SamplesPS(\ell_j) \over
  \ell_i-\ell_j} \smash{\left(1 -
    {\SamplesPS(\ell_i)-\SamplesPS(\ell_j) \over
      \ell_i-\ell_j+1}\right)}$ is the estimated variance for this
stratum.

We call this algorithm \emph{\PropDynPgm} (for dynamic programming for
stratification with proportional allocation).  With analysis similar
to \StraDynPgm\ (except here we only run the dynamic programming
procedure once), we see that the running time of \PropDynPgm\ is
$O(N\log m + Hm^2\log^2N)$, where the two terms can be attributed to
precomputation and dynamic programming, respectively.  The dynamic
programming procedure finds the optimum stratification whose
boundaries are restricted to $B$, which still yields a good
approximation ratio of $2$.

The proof of the next theorem can be found in Appendix~\ref{apndxSec4}.
\begin{theorem}\label{thm:PropDynPgm}
  Given an ordered set \Objs\ of $N$ objects and a sampled subset
  \SamplesI\ of $m$ objects, let $v^*$ denote the minimum value of
  estimated variance defined in~\eqref{eq:stratified-obj:prop}
  achievable using $n$ samples under stratified sampling with
  proportional allocation over $H$ strata.  \PropDynPgm\ runs in
  $O(N\log m + Hm^2\log^2N)$ time and finds a stratification resulting
  in estimated variance $v \leq 2v^*$.
\end{theorem}

As with \StraLogBdr\ and \StraDynPgm, we can improve \PropDynPgm's
approximation ratio at the expense of its running time, by considering
candidate boundary indices that are powers of $(1+\epsilon)$ (instead
of $2$) away from the indices of sampled objects.  The resulting
approximation ratio would become $(1+\epsilon)$ and running time
$O(N\log m + H\smash{m^2 \over \epsilon^2}\log^2 N)$.


\revm{
\section{Experiments}
\label{sec:expts}

Most of our experiments are based on three scenarios, each with its
own real-world dataset and counting query template:
\begin{description}
\item[(\textbf{\emph{Sports}})] The data contains yearly performance
  statistics for players in the Major League Baseball.  We focus on
  pitching statistics, which exclude a portion of the players.  We
  consider the $k$-skyband size query in Example~\ref{ex:k-skyband},
  where each point is a player-year combination (there are about
  $47{,}000$ of them), and \sql{x} and \sql{y} refer to runs and home
  runs.
\item[(\textbf{\emph{Neighbors}})] The data comes from KDD Cup 1999,
  where the goal was to learn a predictive model that could
  distinguish legitimate and illegitimate (intrusion attacks)
  connections to a machine.  The original dataset contains $4.9$
  million records with 41 features and a binary label.  We removed
  many sparse rows, resulting in $73{,}000$ points.  We consider the
  query in Example~\ref{ex:k-distance} that counts points with few
  neighbors.
\item[(\textbf{\emph{Text}})] We consider the relevant document count
  query in Example~\ref{ex:doc-count}.  Since we do not want to
  manually evaluate the predicate ourselves in experiments, we use the
  \emph{LSHTC} dataset~\cite{partalas2015lshtc}, which provides
  ground-truth labels (Wikipedia categorization) for 2.4M documents
  from Wikipedia.  The same dataset was used
  in~\cite{lu_accelerating_2018}.  In our experiments, each algorithm
  is charged a cost for revealing the true label, which in practice
  would be expensive.
\end{description}
To experiment with different selectivities of the predicate \Pred, we
adjust query parameter settings ($k$ for \emph{Sports}; $k$ and $d$
for \emph{Neighbors}; $k$ for \emph{Text}).  We also create synthetic
datasets based on \emph{Sports} to study how data distributions affect
learned models and the performance of various algorithms; for details
see Section~\ref{sec:expts:predictability}.


We compare the following algorithms:
\begin{itemize}
\item Sampling-based (Section~\ref{sec:prelim:sample}): simple random
  sampling (\SRS) and stratified sampling (\SSP, with proportional
  allocation, and \SSN, with Neyman allocation in two stages).  For
  stratified sampling (which applies to \emph{Neighbors} and
  \emph{Sports} but not to \emph{Text}), we use attributes \sql{x} and
  \sql{y} as surrogates; each stratum is a rectangle in the 2d
  \sql{x}-\sql{y} space.  Unless otherwise specified, we stratify
  using a uniform $\sqrt{H}\times\sqrt{H}$ grid over the ranges of
  \sql{x} and \sql{y} values in the dataset.  By default $H=4$.
\item Learning-based (Section~\ref{sec:prelim:learn}): quantification
  learning (\QLCC, without adjustment, and \QLAC, with adjustment).
\item Learning with sampling-based correction
  (Section~\ref{sec:prelim:correct}): \QLSC.
\item Learning-to-sample (Section~\ref{sec:approach}): learned
  weighted sampling (\LWS) and learned stratified sampling (\LSS).
  Unless otherwise specified, for \LSS\ we implement a simplified
  version of \StraLogBdr, which considers candidate boundaries that
  map to equally spaced ticks over $[0,1]$ (the range of $g$ scores).
  By default, $H=4$ and the spacing between candidate boundaries is
  $0.05$; for the distributions of $g$ scores that arise in practice,
  these boundaries already provide fine enough resolution for $H=4$,
  so more sophisticated choices of candidates in \StraLogBdr\ are not
  needed.
\end{itemize}
For learning-based and learn-to-sample algorithms, we use standard
implementations of classifiers from \sql{scikit-learn}.  For
\emph{Neighbors} and \emph{Sports}, we experiment with \textsf{kNN}
($k$-nearest neighbors, where $k$ is not to be confused with our query
parameter), \textsf{RF} (random forests), and \textsf{NN} (a simple
two-layer neural network); by default, we use \textsf{RF} with $100$
estimators.  For \emph{Text}, we use a naive Bayes classifier with
standard full-text features.
For \QLSC, \LWS, and \LSS, by default we devote 25\% of their allotted
samples to training (and including design, if applicable).

%
Since the estimates of result counts are uncertain, for each
experimental setting, we run each algorithm $100$ times, and record
the distribution of estimates it produces.  Recall that unlike
sampling-based and learn-to-sample algorithms, those based on learning
alone provide no accuracy guarantees by themselves.  Nonetheless, the
distributions of estimates they produce allow us to evaluate their
accuracy empirically.  When appropriate, we show distributions using
violin plots\footnote{A violin plot shows the probability density at
  different values; additionally, a white dot marks the median of all
  data, a thick black line spans the lower and upper quartiles.}.  We
would like our estimates to be unbiased, so ideally the violin plots
would be centered around the actual result count.  Furthermore, we
would like the estimates to have low variance, which means narrower
interquartile ranges as well as shorter and wider plots.  In some
figures, we use MAE (mean absolute error) as a single numeric measure
to quantify and summarize an error distribution, so we can report more
results than violin plots.

For \emph{Neighbors} and \emph{Sports}, while our queries can be
executed directly over a database system, they run slowly even if we
construct all appropriate standard indices and enable the maximum
level of optimization (on PostgreSQL and another commercial system).
To enable faster experiments, we implemented the evaluation of \Pred\
in Python in main memory.  Since our experiments specify sampling
budgets in terms of numbers (or percentages) of samples, our results
are platform-neutral and easy to translate into time savings on
different underlying platforms.  The overhead of learning, as we will
show later with experiments, is small compared to the cost of labeling
samples (evaluating \Pred), even for the in-memory Python
implementation; the overhead will be even smaller in the SQL setting.

\subsection{Overall Comparison with Real Datasets}
\label{sec:expts:overall}

We begin with experiments that compare various algorithms using the
three scenarios with real datasets, \emph{Neighbors},
\emph{Sports}. and \emph{Text}.  Both \LSS\ and \LWS\ used a random
forest classifier with estimators and a 25\%:75\% training:sampling
split.  Figure~\ref{fig:overall-vary-size} compares the MAE of various
algorithms when we vary the result size (via query parameters) while
keeping the sample size fixed.  Figure~\ref{fig:overall-vary-sample}
compares the MAE of various algorithms when we vary the sample size
while keeping the result size fixed.

\begin{figure*}[ht]
  \begin{subfigure}[b]{0.33\textwidth}
    \caption{\emph{Neighbors}}
    \includegraphics[width=.99\textwidth]{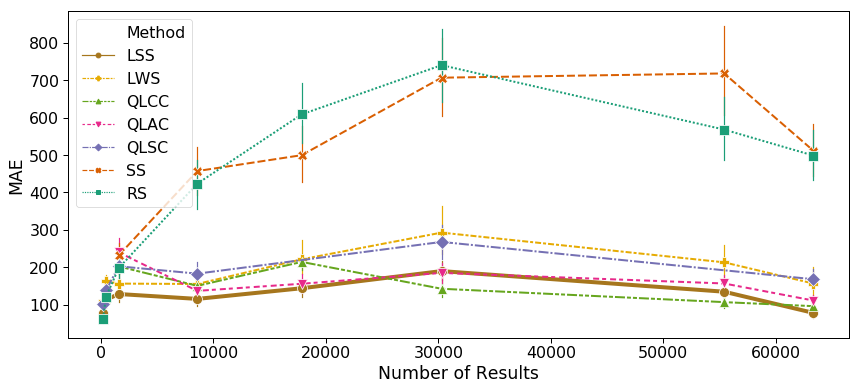}
    \label{fig:overall-vary-size:neighbors}
  \end{subfigure}
  \begin{subfigure}[b]{0.33\textwidth}
    \caption{\emph{Sports}}
    \includegraphics[width=.99\textwidth]{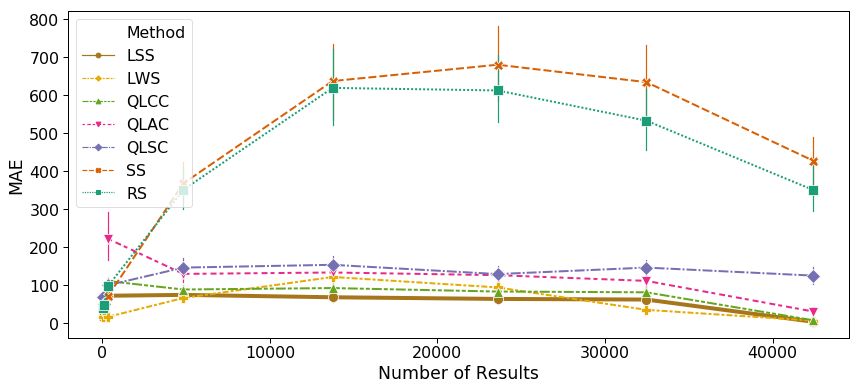}
    \label{fig:overall-vary-size:skyband}%
  \end{subfigure}
  \begin{subfigure}[b]{0.33\textwidth}
    \caption{\emph{Text}}
    \includegraphics[width=.99\textwidth]{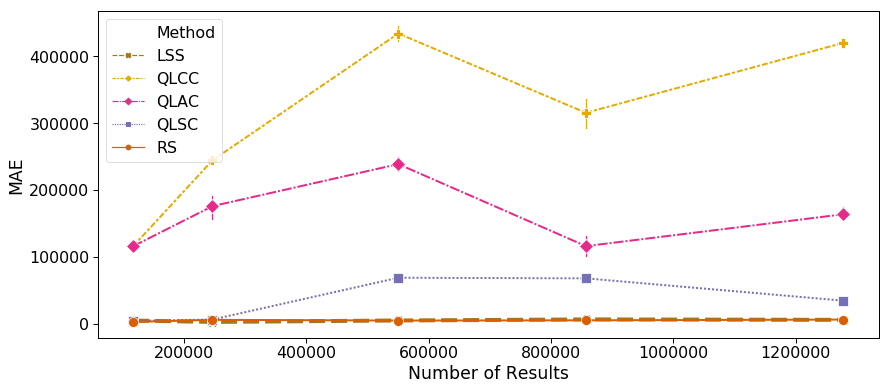}
    \label{fig:overall-vary-size:text}%
  \end{subfigure}
  \vspace*{-5ex}
  \caption{\revm{Mean absolute error comparison when varying result size;
    sample size fixed at $2\%$.}}
  \label{fig:overall-vary-size}
\end{figure*}

\begin{figure*}[ht]
  \begin{subfigure}[b]{0.33\textwidth}
    \caption{\emph{Neighbors}}
    \includegraphics[width=.99\textwidth]{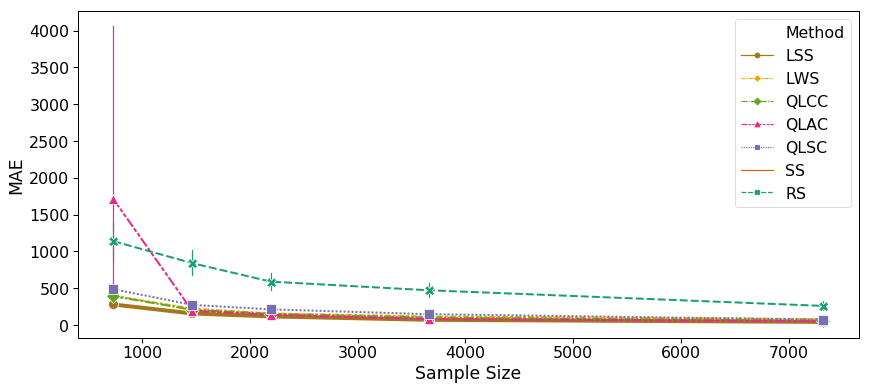}
    \label{fig:overall-vary-sample:distance}
  \end{subfigure}
  \begin{subfigure}[b]{0.33\textwidth}
    \caption{\emph{Sports}}
    \includegraphics[width=.99\textwidth]{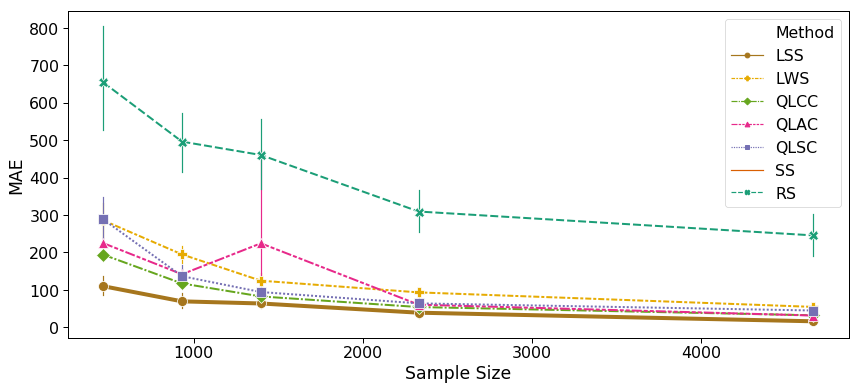}
    \label{fig:overall-vary-sample:skyband}%
  \end{subfigure}
  \begin{subfigure}[b]{0.33\textwidth}
    \caption{\emph{Text}}
    \includegraphics[width=.99\textwidth]{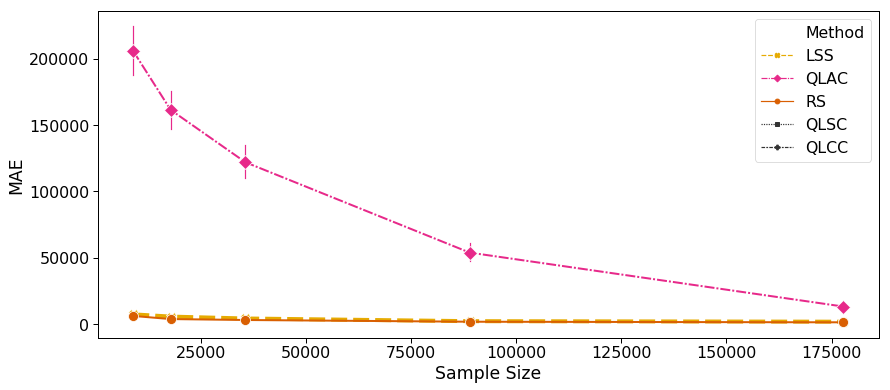}
    \label{fig:overall-vary-sample:text}%
  \end{subfigure}
  \vspace*{-5ex}
  \caption{\revm{Mean absolute error comparison when varying sample size.}}
  \label{fig:overall-vary-sample}
\end{figure*}

As it turns out, the learned classifier performs pretty well for
\emph{Neighbors} and \emph{Sports}, but pretty poorly for \emph{Text},
leading to very different results.  We shall focus on \emph{Neighbors}
and \emph{Sports} first.  F1 scores for the learned classifiers
average higher than $0.8$ in these scenarios (with small result sizes
being more difficult).  We make several observations.  \textbf{First},
learning-based methods are very competitive here thanks to high
classifier quality.  In fact, \QLCC\ sometimes even delivers the
smallest errors even without any adjustment or correction.  But to
keep things in perspective, \QLCC\ and \QLAC\ do not provide any
guarantees; once \QLSC\ uses sampling to provide correction and
guarantees, MAE actually takes a small hit because of the extra
overhead.  \textbf{Second}, algorithms without any learning component,
namely \SRS\ and \SSP\ are clearly not as competitive here, with much
higher MAE than others.  \textbf{Third}, \LSS\ (highlighted) has
consistently low MAE; it is nearly always the leader or not far from
the leader, and bear in mind that it offers statistical guarantees,
which \QLCC\ does not.  \LSS\ also consistently leads \QLSC\ by a good
margin.  \textbf{Fourth}, the comparison between \LWS\ and \LSS\ is
difficult, as in some cases \LWS\ leads \LSS.  The quality of the
learned classifier for \emph{Neighbors} and \emph{Sports} is the main
factor here.  To better understand the situation, we take a closer
look at some data points with violin plots showing distributions.

In Figure~\ref{fig:lss_vs_lws}, we get a more detailed sense of the
variability in estimates.  \LSS\ and \LWS\ are consistently no worse
and often better than \SRS\ and \SSP.  Between \LSS\ and \LWS, we make
two observations.  \textbf{First}, when selectivity is low, we expect
all sampling-based methods to have some trouble as the particular
number of positives that come up by chance in each run will have a
large impact on relative error.  For \emph{Sports}, \LWS\ dodged this
issue with a very good classifier that allows it to draw in a very
targeted fashion.  In contrast, \LSS, as it places much less trust in
the learned model compared with \LWS, misses the opportunity.
\textbf{Second}, \LWS\ is not without its own problems.  In
\emph{Neighbors}, where prediction becomes slightly more challenging,
we see \LWS\ underestimating with XS result size; as it turns out, the
classifier at those points happens to generate more false negatives.
In other words, \LWS\ depends far more on model quality than \LSS\
does---it can benefit more, but also can get hurt more.  This effect
will be magnified for the \emph{Text} scenario, which we focus on
next.

\begin{figure*}[ht]
  \centering
  \begin{subfigure}[b]{0.48\textwidth}
    \caption{\emph{Neighbors} (XS/S/L = 2/10/40\%)}
    \includegraphics[width=.99\textwidth]{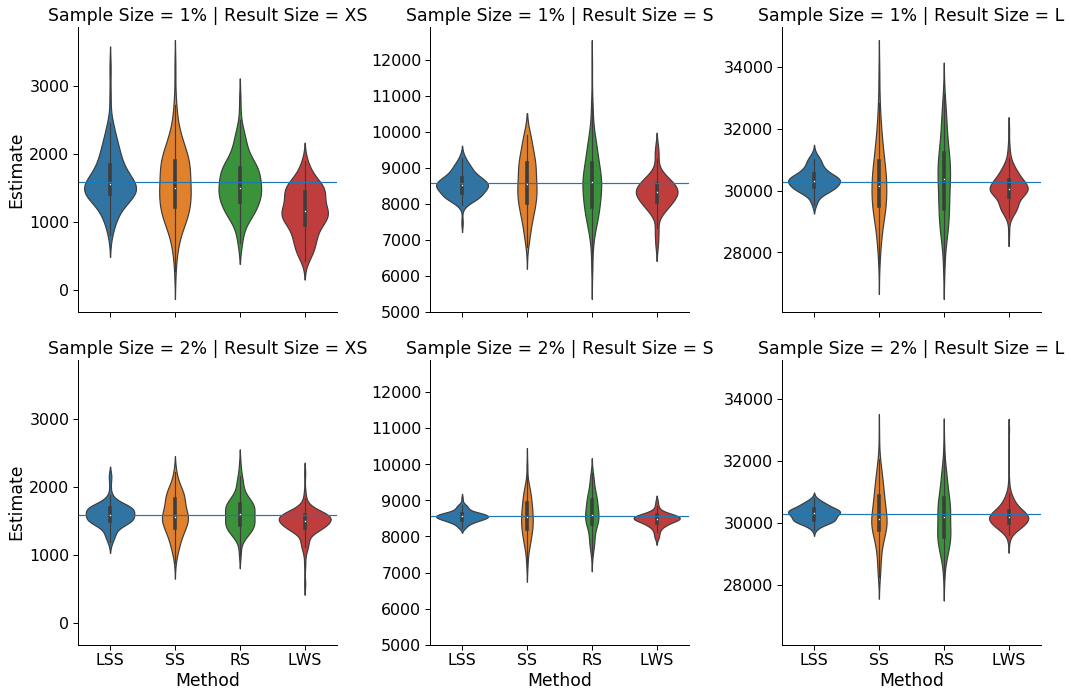}
    \label{fig:lss_vs_lws:distance}
  \end{subfigure}
  \begin{subfigure}[b]{0.48\textwidth}
    \caption{\emph{Sports} (XS/S/L = 1/10/50\%)}
    \includegraphics[width=.99\textwidth]{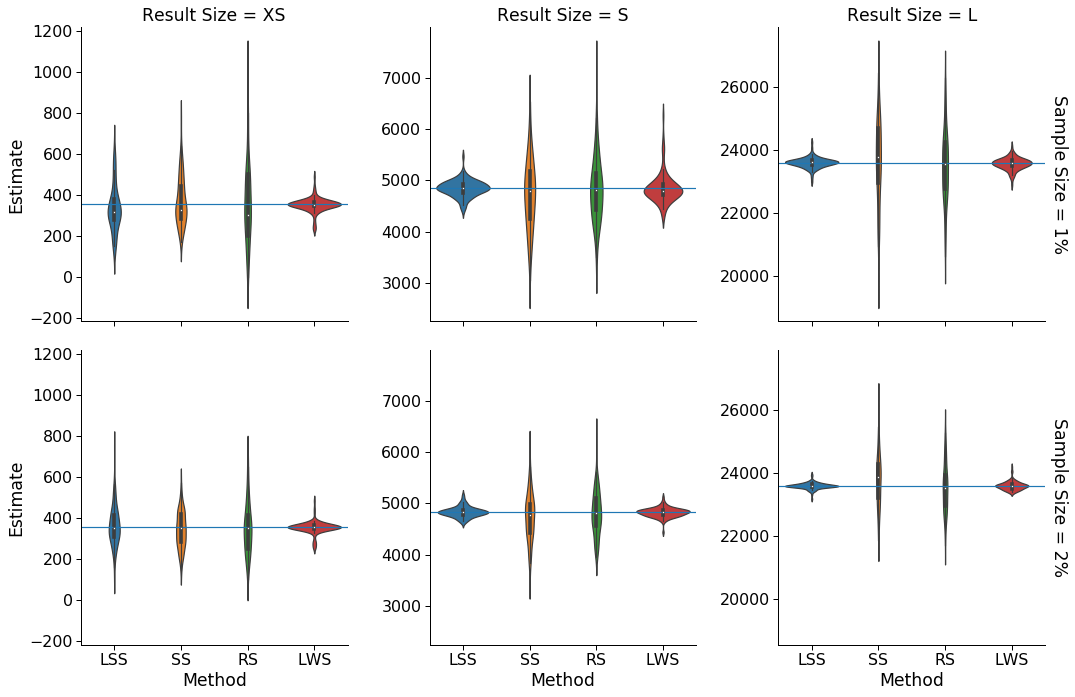}
    \label{fig:lss_vs_lws:skyband}%
  \end{subfigure}
  \caption{\revm{Distributions of estimates.  Each row has a different
    sample size (1\%, 2\%), and each column has a different result
    size.}}
  \label{fig:lss_vs_lws}
\end{figure*}

The \emph{Text} tells a completely different story.  In this case,
classification is hard.  Therefore, \QLCC, \QLAC, and \QLSC\ fare very
poorly here, because their performance is too dependent on starting
point produced by \QLCC.  Correction is also difficult.  From one
representative run (with $857$k resize size), true TPR and FPR are
$.53$ and $.85$, while the estimated TPR and FPR are $.35$ and $.95$.
Even with sampling-based correction, \QLSC\ still underperforms other
algorithms.  In contrast, \SRS, which does not use learning, actually
shines here.  Finally, \LSS\ tracks \SRS\ closely.  It actually
underperforms \SRS\ a bit, which is understandable because learning
phase is essentially not that useful, wasting 25\% of the samples.
However, the impact on the sampling design is limited.  Closer
examination reveals that it basically degenerates to \SRS\ for the
remaining 75\% of the samples.  This experiment highlights the
sensitivity of \QLCC, \QLAC, and \QLSC\ toward poor models, as well as
the resiliency of \LSS\ against poor models.


\subsection{Comparison with Synthetic Datasets}
\label{sec:expts:predictability}

Results in Section~\ref{sec:expts:overall} show just three data points
along the spectrum of classifier quality: \emph{Neighbors} and
\emph{Sports} have good classifiers but \emph{Text} has a bad one.
What happens in between?  To understand how different algorithms are
affected by varying degrees of difficulty in using a learned model to
approximate a predicate, we design our next set of experiments by
injecting additional ``noise'' into the \emph{Sports} scenario to
adjust the difficulty of classification.  Recall from
Example~\ref{ex:k-skyband} that for each object $o$, we compute a
count subquery with $o.\sql{x}$ and $o.\sql{y}$, and compare the
resulting count, say $c$, with $k$.  Now, we create an additional
``noise'' table keyed on distinct $(\sql{x}, \sql{y})$ values, where
each $(\sql{x}, \sql{y})$ is associated with a noise count drawn
randomly from another distribution.  Instead of comparing $c$ with
$k$, we use another subquery to look up the noise count $c'$ for
$(o.\sql{x}, o.\sql{y})$, and have the predicate combine the original
and noise counts into $(1-\alpha) c + \alpha c'$ to compare with $k$.
By adjusting $\alpha \in [0,1]$, we control how much noise contributes
to the outcome of the predicate: $\alpha = 0$ corresponds to the
original \emph{Sports} scenario, where we know we can learn a good
model; $\alpha = 1$ means the predicate is simply comparing
independent random noise, which is mostly challenging to predict.

We experiment with two noise distributions.  One is a Gaussian with
standard deviation of $1$ truncated and discretized.  The other is
derived from a Zipf distribution with parameter $s$, where each draw
is used to index into a randomly permuted array of possible noise
counts derived from the real count values; large $s$ means some
(random) noise count will be far more popular than others.

We compare \SRS, \QLSC, and \LSS, representing sampling-based,
learning-based (but with sampling-based correction), and
learn-to-sample algorithms, respectively.
Figure~\ref{fig:synthetic_gaussian} shows how they compare in terms of
MAE when we vary $\alpha$ for synthetic datasets generated using
Gaussian noise.  Note that when $\alpha$ increases, the result size
tends to decrease (but it is random depending on the particular
dataset being generated), so MAE for \SRS\ tends to decrease
accordingly, although its relative error actually increases. The main
observation from this figure is that when $\alpha$ is small, good
model qualities make \LSS\ and \QLSC\ outperform \SRS.  However, as
$\alpha$ increases, model quality starts to take a toll on \LSS\ and
\QLSC.  Nonetheless, \LSS\ consistently outperforms \QLSC, and it is
not too far behind \SRS\ even when the predicate outcome is almost
completely dictated by noise.  Upon closer examination, we see that
when $\alpha=1$, \LSS\ basically degenerates to random sampling in the
sampling phase, and it is not surprising that it is slightly worse
than \SRS\ because it has wasted 25\% of its samples on learning.

Figure~\ref{fig:synthetic_zipf} shows how the three algorithms compare
when we vary the Zipf parameter for synthetic datasets generated using
Zipf noise.  The results can be difficult to interpret because of the
variability in each particular instance of the randomly generated
dataset, and the fact that skewness does not necessarily make
classification harder.  However, once we overlay the quality (F1
score) of learned classifier for \QLSC\ and \LSS, a clear pattern
emerges: model quality clearly influences the performance of methods
that use learning, but \LSS\ is far more resilient than \QLSC\
(consider $s=7$, for example).  Again, \LSS\ is the most consistent
performer among all three---it is not far from \SRS\ when the model is
very poor, and it is not far from \QLSC\ when the model is very good.


}
\subsection{Running Time and Overhead}
\label{sec:expts:runningtime}

\begin{figure*}[t]
  \begin{minipage}[t]{0.32\textwidth}
  \centering
  \includegraphics[width=.8\linewidth]{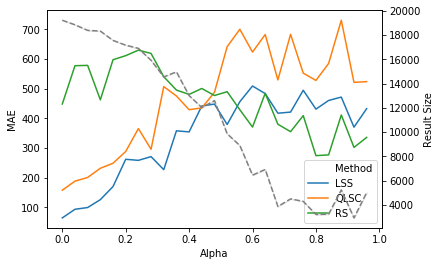}
  \caption{\label{fig:synthetic_gaussian}\revm{Varying $\alpha$; synthetic
    datasets with Gaussian noise; $k=15000$.  Grey dashed line shows the
    result size (scale on right).}}
  \end{minipage}
  \hfill
  \begin{minipage}[t]{0.32\textwidth}
  \centering
  \includegraphics[width=.8\linewidth]{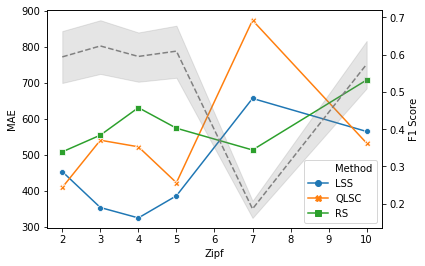}
  \caption{\label{fig:synthetic_zipf}\revm{Varying Zipf parameter $s$;
      synthetic datasets with Zipf noise; $\alpha=0.6$; $k=15000$.
      Grey dashed line/band show the F1 scores of the classifier
      (scale on right).}}
  \end{minipage}
  \hfill
  \begin{minipage}[t]{0.32\textwidth}
  \centering
  \includegraphics[width=.8\linewidth]{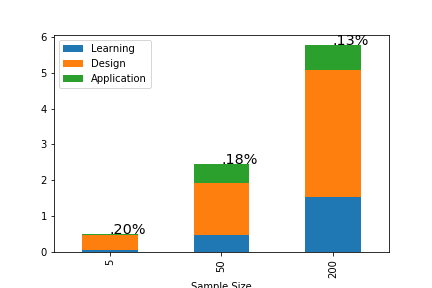}
  \caption{\label{fig:execution_time_overhead}Execution time overhead
    (in seconds) vs.\ sample size (in thousands) for \LSS, broken down
    by sources of overhead.}
  \end{minipage}
\end{figure*}

Before making a closer examination of \LSS, we take a brief look at
the running times of our approaches. Both \LWS\ and QL methods (\QLAC,
\QLCC, \QLSC) are simpler than \LSS, which has more overhead in
stratification.  Thus, we focus on \LSS.  In
Figure~\ref{fig:execution_time_overhead}, we plot the overhead added
by using \LSS\ when compared with \SRS. There are three distinct
sources of overhead in \LSS: \emph{Learning} represents the time to
train the classifier; \emph{Design} includes the time to compute the
optimal stratified sampling scheme; \emph{Application} accounts for
the overhead in applying the chosen scheme, which involves picking
objects from their associated strata.  (Note that we already charge
the samples used by \LSS\ for learning and sampling design towards the
total number of samples, which is set to be the same when comparing
with other approaches.)  In Figure~\ref{fig:execution_time_overhead},
we also list the fraction of overall running time consumed by overhead
at the top of each bar.  Note these are miniscule (below $0.2\%$)
compared with the overall cost, dominated by the predicate evaluation
over samples.  Such a low overhead implies that if we give simpler
approaches such as \SRS\ additional samples to account for the
overhead of \LSS, the number of additional samples would be too low to
make any difference.

\subsection{Closer Looks at LSS}
\label{sec:expts:lss}

Next, we test a variety of facets involved in \LSS: how strata are
laid out, the number of strata, allocation of samples for
learning/design vs.\ estimation, and how the underlying classifier
affects final estimation quality.

\begin{figure*}[ht]
  \centering
  \begin{subfigure}[b]{0.48\textwidth}
    \caption{\emph{Neighbors} (XS/S/L = 2/10/40\%)}
    \includegraphics[width=.99\textwidth]{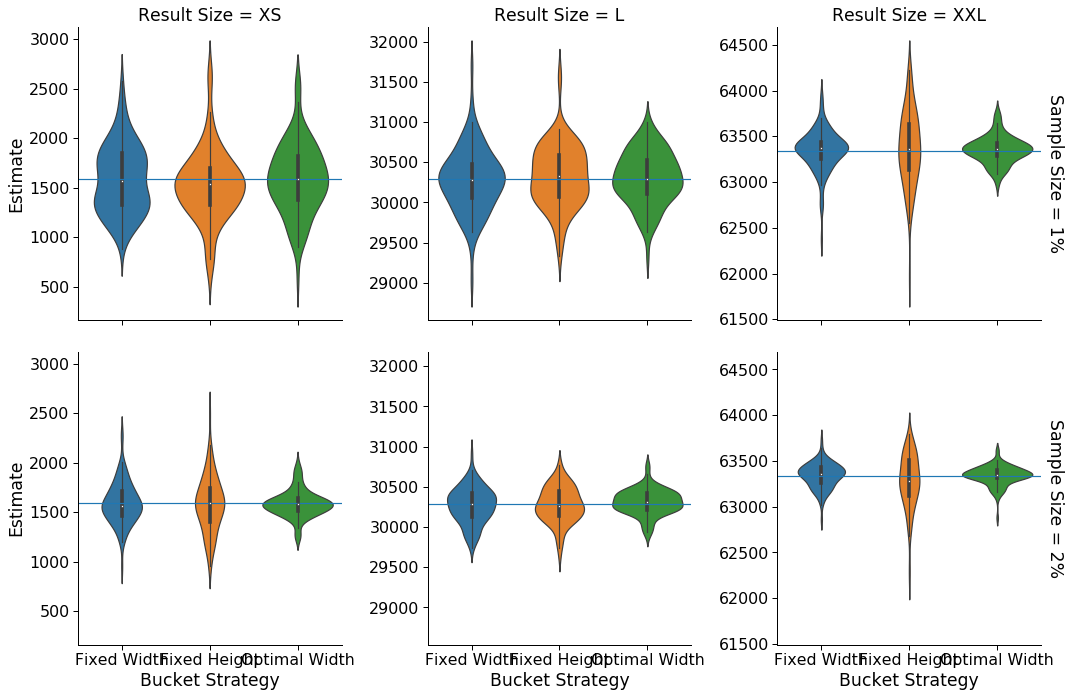}
  \end{subfigure}
  \begin{subfigure}[b]{0.48\textwidth}
    \caption{\emph{Sports} (XS/S/L = 1/10/50\%)}
    \includegraphics[width=.99\textwidth]{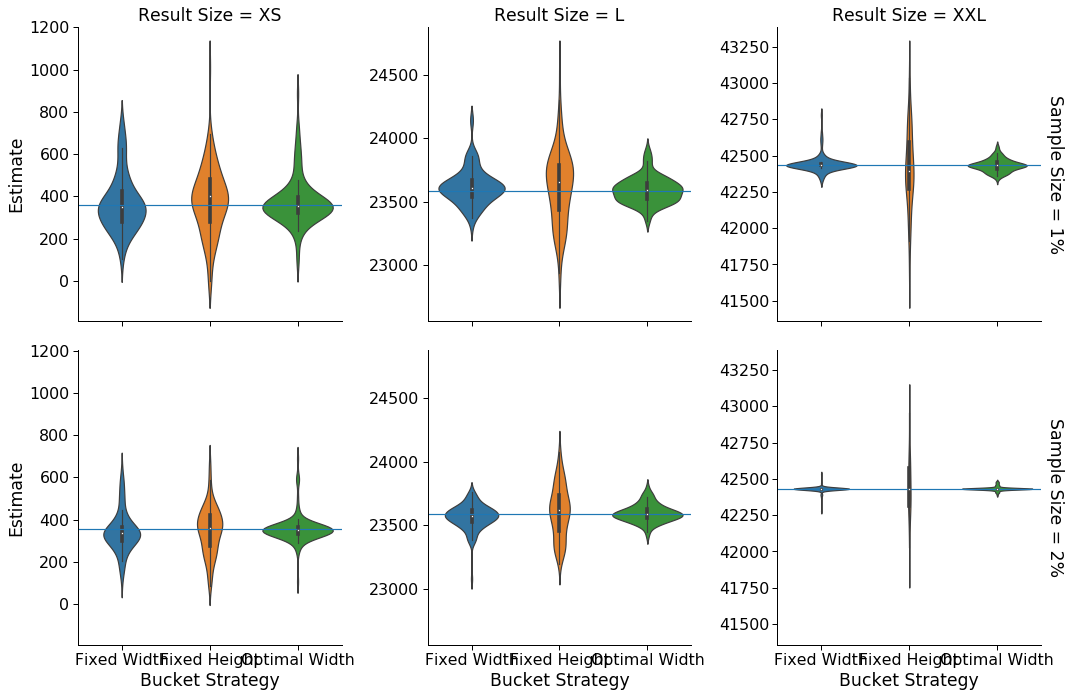}
  \end{subfigure}
  \caption{Effect of stratification strategy on \LSS\ estimation
    quality. Each row represents a different sample size (1\%, 2\%),
    and each column represents a change of parameters resulting in a
    different result set size (XS, S, L).}
  \label{fig:strata_layout}
\end{figure*}

\paragraph{Strata Layout Strategy}
First, we study the impact of stratification strategy on \LSS.
Instead of using more sophisticated algorithms to look for optimal
bucket boundaries (\emph{optimal-width}), what if we use simpler
strategies?  In particular, \emph{fixed-width} simply make all strata
equal in width; \emph{fixed-height} simply ensures that all strata
contain the same number of objects.  Figure~\ref{fig:strata_layout}
shows the results, using 4 strata.  It is no surprise that
\emph{fixed-height} produces poor results for stratified sampling, as
each strata may be force to contain a mixture of labels; in
particular, for skewed datasets where one label occurs more often (XS
and XXL), \emph{fixed-height} has much higher variance in its
estimates.  \emph{Fixed-width} fares better, but our
\emph{optimal-width} (which \LSS\ uses by default) makes further
gains---its interquartile range (IQR) is generally lower than the two
simpler approaches.

\begin{figure*}[ht]
  \centering
  \begin{subfigure}[b]{0.49\textwidth}
    \caption{\emph{Neighbors} (XS/S/L = 2/10/40\%)}
    \includegraphics[width=.99\textwidth]{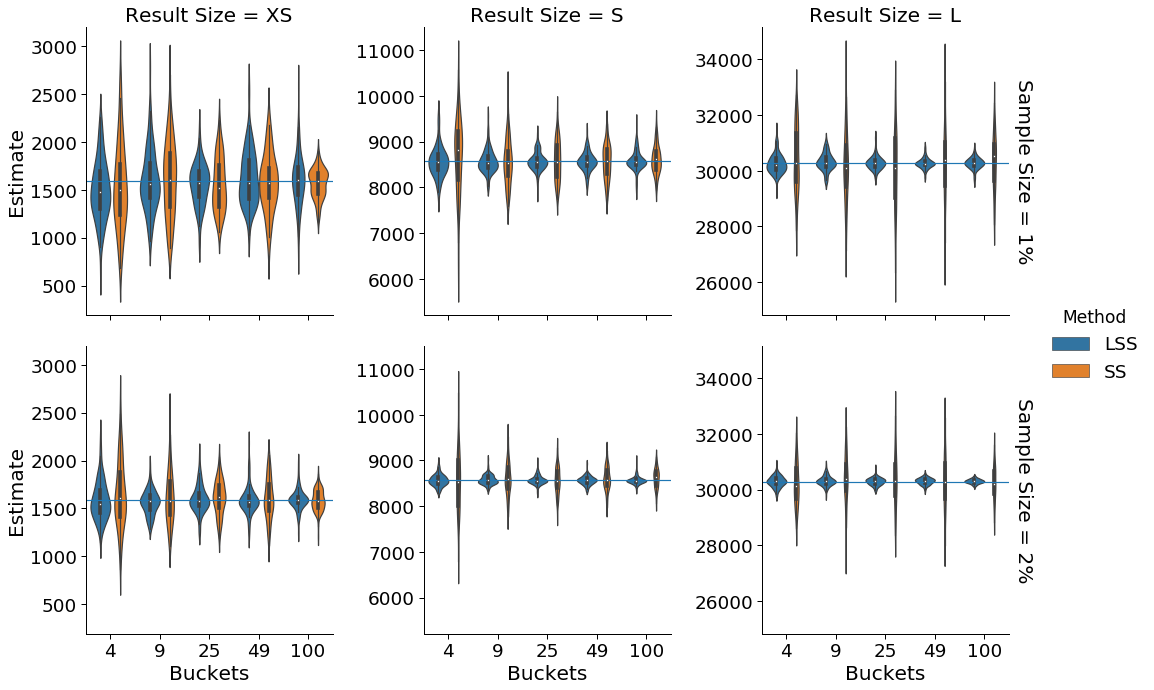}
    \label{fig:distance-number-bins-ss}
  \end{subfigure}
  \begin{subfigure}[b]{0.49\textwidth}
    \caption{\emph{Sports} (XS/S/L = 1/10/50\%)}
    \includegraphics[width=.99\textwidth]{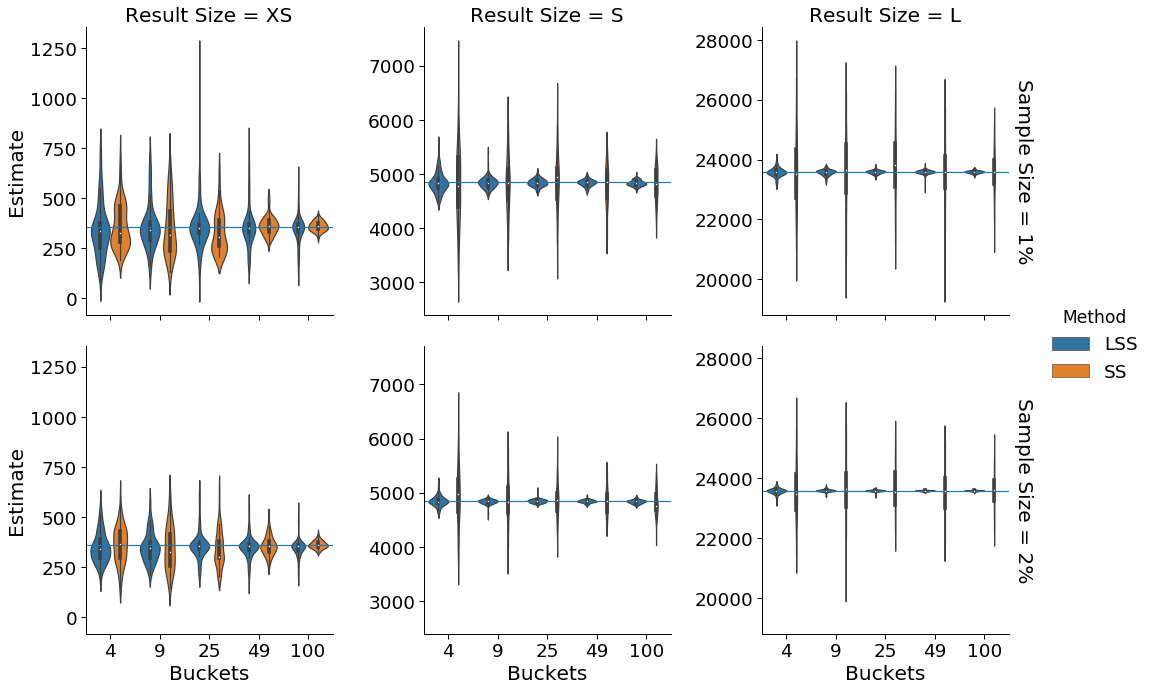}
    \label{fig:kskyband-number-bins-ss}%
  \end{subfigure}
  \caption{Comparison of \LSS\ and \SSP\ across varying number of
    strata. Each row represents a different sample size (1\%, 2\%),
    and each column represents a change of parameters resulting in a
    different result set size (XS, S, L).}
  \label{fig:lss:ssp:num_strata}
\end{figure*}

\paragraph{Number of Strata}
In this experiment, we investigate the effect of the number of strata
on estimation quality when using \LSS\ and \SSP, both of which use
stratified sampling.  We vary the number of strata with 4, 9, 25, 49,
and 100 strata available.  The results are summarized in
Figure~\ref{fig:lss:ssp:num_strata}.  Overall, as expected, increasing
the number of strata tends to improve estimation quality, but not
substantially so.  Here, with XS result size and a large number of
strata, \SSP\ becomes competitive against \LSS.  The reason is that
with superfine uniform gridding of the $\sql{x}$-$\sql{y}$ space, the
few positive objects eventually concentrate into a few strata, making
\SSP\ effective; in comparison, \LSS\ may occasionally produce an
outlier estimate, even though its overall variance is still
competitive.  Aside from these few extreme settings, however, \LSS\
generally outperforms \SSP, and often by significant margins as shown
in Figure~\ref{fig:lss:ssp:num_strata}.

\begin{figure*}[ht]
  \centering
  \begin{subfigure}[b]{0.48\textwidth}
    \caption{\emph{Neighbors} (XS/S/L = 2/10/40\%)}
    \includegraphics[width=.99\textwidth]{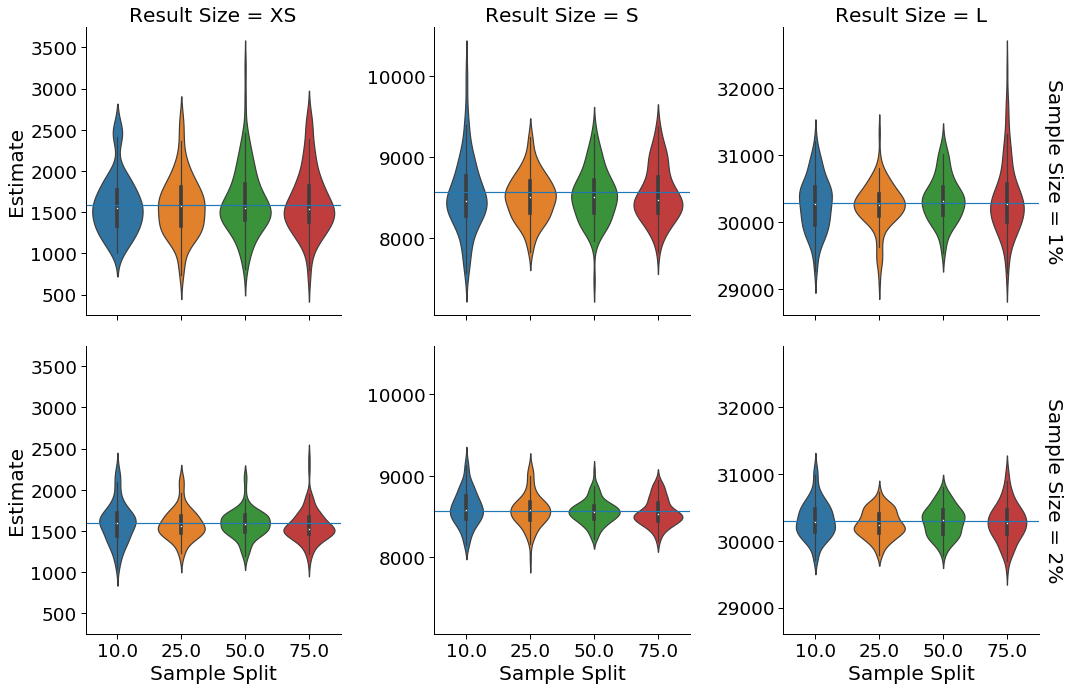}
    \label{fig:distance-sample-split}
  \end{subfigure}
  \begin{subfigure}[b]{0.48\textwidth}
    \caption{\emph{Sports} (XS/S/L = 1/10/50\%)}
    \includegraphics[width=.99\textwidth]{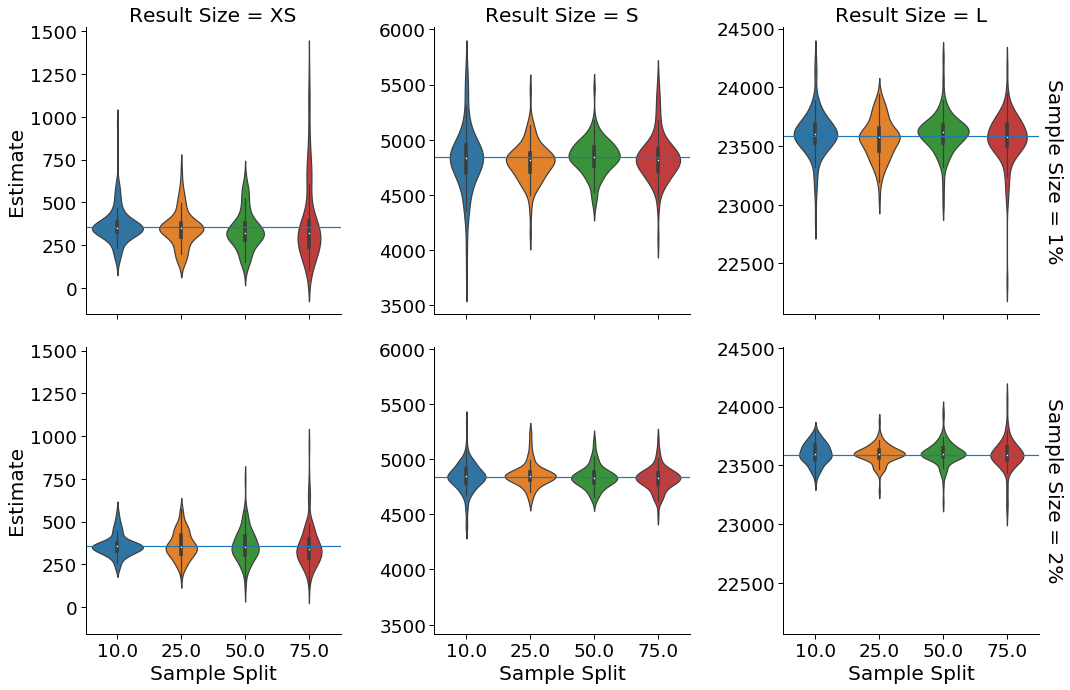}
    \label{fig:kskyband-sample-split}%
  \end{subfigure}
  \caption{\LSS\ as the percentage of samples used for learning/design
    (instead of producing the result estimate) varies. Each row
    represents a different sample size (1\%, 2\%), and each column
    represents a change of parameters resulting in a different result
    set size (XS, S, L).}
  \label{fig:sample_split}
\end{figure*}

\paragraph{Sample Split}
Next, we test the effect of sample allocation on the quality of
estimates produced by \LSS.  We vary the percentage of samples
allocated to classifier training and sampling design from 10\%, 25\%,
50\%, to 75\%.  A 10\% split means 10\% of the total samples are
devoted to learning and design, while the rest (90\%) of the samples
are used to produce the result estimate.  We fix the number of strata
at 4.  Figure~\ref{fig:sample_split} summarizes the results.  We see
that at 75\%, too few samples are devoted to estimation, so the result
quality tends to suffer.  Conversely, at 10\%, too few samples are
devoted to learning and design, and the result quality may also
suffer.  Both middle proportions (25\% and 50\%) consistently produce
the most reliable estimates with lowest IQR's and fewer outliers.

\begin{figure*}[ht]
  \centering
  \begin{subfigure}[b]{0.48\textwidth}
    \caption{\emph{Neighbors} (XS/S/L = 2/10/40\%)}
    \includegraphics[width=.99\textwidth]{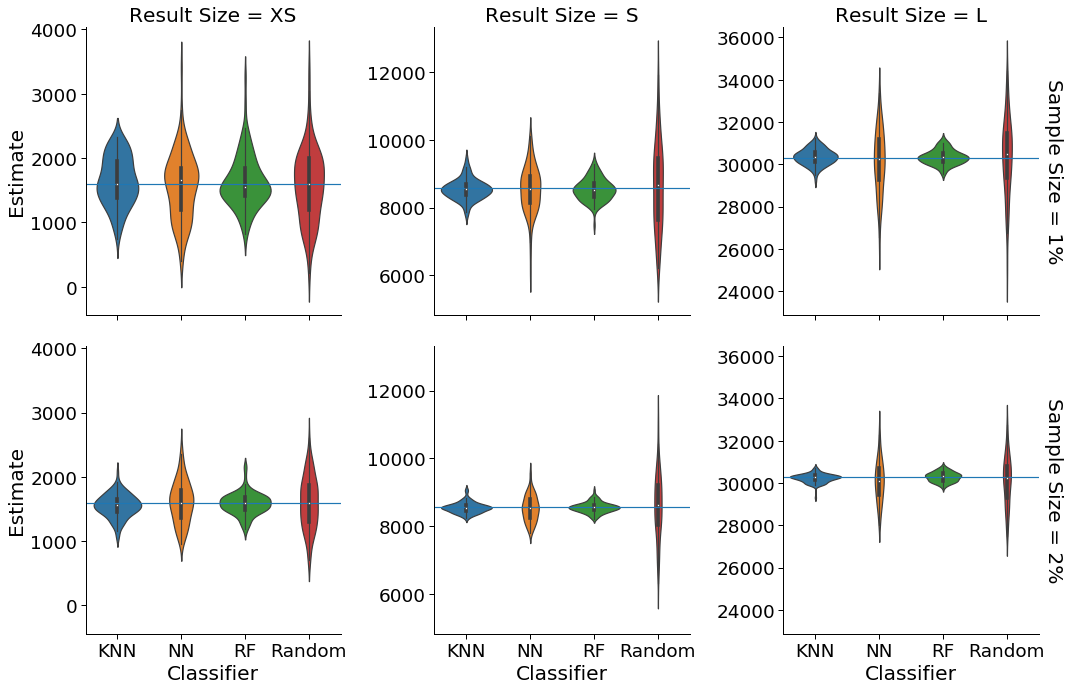}
    \label{fig:distance-number-bins}
  \end{subfigure}
  \begin{subfigure}[b]{0.48\textwidth}
    \caption{\emph{Sports} (XS/S/L = 1/10/50\%)}
    \includegraphics[width=.99\textwidth]{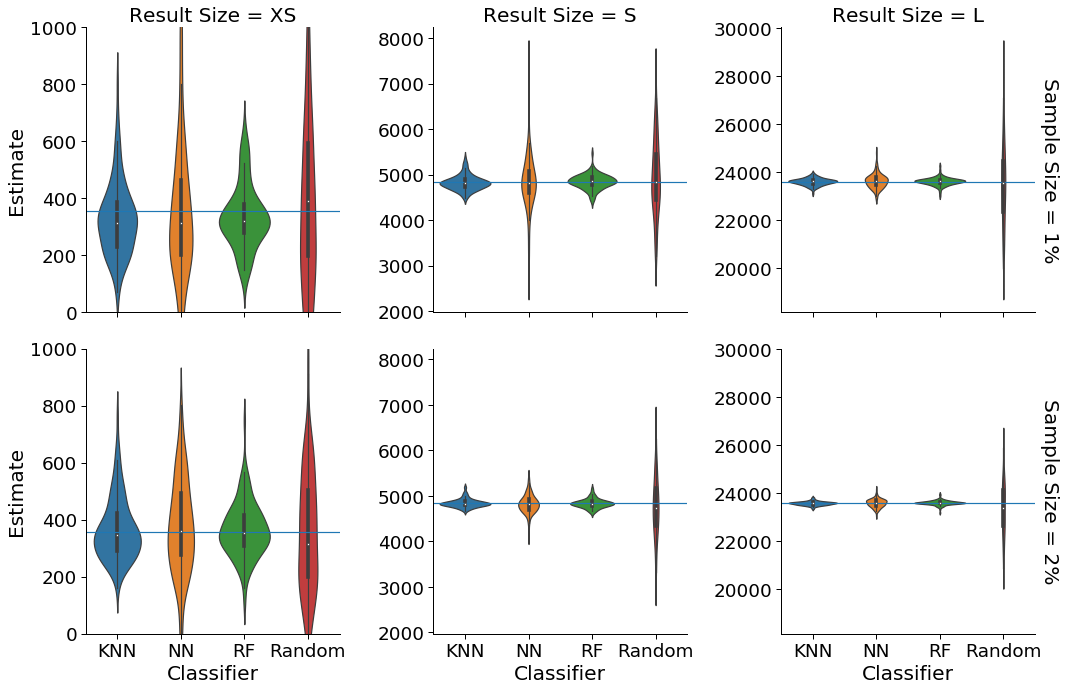}
    \label{fig:kskyband-number-bins}%
  \end{subfigure}
  \caption{\LSS\ under different classifiers. Each row represents a
    different sample size (1\%, 2\%), and each column represents a
    change of parameters resulting in a different result set size (XS,
    S, L).}
  \label{fig:classifiers}
\end{figure*}

\paragraph{Choice of Classifier}
As \LSS\ is driven by the scores produced by a classifier, it is
naturally dependent on the classifier itself.  We tested \LSS\ with
four classifiers: $k$-Nearest Neighbors (KNN, with $k=3$), simple
two-layer neural network (NN, with $5$ nodes per layer), random forest
(RF, with 100 estimators), and a dummy classifier (Random) that
assigns arbitrary random scores to objects.  Random can be viewed as a
worst case scenario for \LSS\ as the desired effect of stratification
(producing homogeneous strata) is completely lost.  Across
classifiers, we use 25\% of the samples for learning and sampling
design, and there are $4$ strata.  As we can see from the results in
Figure~\ref{fig:classifiers}, consistent with intuition, a classifier
that performs better than Random produces better estimates.  On the
other hand, even if a classifier performs poorly (such as Random),
\LSS\ still produces reasonable estimates.

\section{Related Work}
\label{sec:related}

\paragraph{Sampling for Approximate Query Processing (AQP)}
Sampling is a fundamental problem in databases and has been studied
over more than three decades~\cite{olken_simple_1986, Olken93thesis,
  chaudhuri_random_1999}.  Random samples are one of the key types of
\emph{synopses}~\cite{fntdb12-CormodeGarofalakisEtAl-synopses_massive_data}
frequently used for AQP.  Sampling for complex queries has been a
long-standing challenge.  In particular, sampling over joins is
non-trivial, because simply joining independent samples of
participating tables is ineffective~\cite{Olken93thesis,
  chaudhuri_random_1999}.  This problem has received much attention
over the years, with representative works such as \emph{ripple
  join}~\cite{haas_ripple_1999},
\emph{wander-join}~\cite{li_wander_2016}, and more recently, sampling
multi-way acyclic and cyclic joins~\cite{Zhao+2018}.  \reva{The focus
  of our work is on counting queries with complex predicates.  Even
  though our predicates can include joins as discussed in
  Section~\ref{sec:problem}, our techniques differ because of
  different problem assumptions.  First, some work on sampling over
  joins, e.g.,~\cite{chaudhuri_random_1999}, aims at producing a
  random sample of the result tuples (i.e., answering a reporting query),
  while we aim at estimating the result count (i.e., answering a counting query).
  Second, to make our approach general, we adopt a
  rather simple evaluation model, where sampling a candidate object
  $o$ to be counted involves evaluating $\Pred(o)$ exactly, without
  additional sampling or approximation.  In contrast, much of the work
  on sampling over joins assumes specific forms of join predicates or
  availability of indexes to avoid enumerating all join results for
  $o$.  On the other hand, all queries in our experiments are too
  complex for these approaches to handle, because these queries use
  constructs such as self-joins, complex non-equality join predicates,
  subqueries containing \sql{GROUP} \sql{BY} and \sql{HAVING}, as well
  as UDFs.}

\reva{\emph{BlinkDB}~\cite{Agarwal:2013:BQB:2465351.2465355} and
  \emph{VerdictDB}~\cite{DBLP:conf/sigmod/ParkMSW18} are examples of
  recent AQP systems aimed at supporting approximate processing of
  general, ad hoc queries.  While these systems deliver very fast
  response time thanks to optimizations such as precomputation and
  parallelization, handling the full complexity of SQL remains
  challenging.  For instance, \emph{VerdictDB} does not support
  self-joins out of the box; our best attempt at adapting the query in
  Example~\ref{ex:k-skyband} to run on it resulted in poor estimates
  compared with other approaches we experimented with in
  Section~\ref{sec:expts}.}

A number of papers are related to ours in the use of sampling.
\cite{DBLP:conf/edbt/NguyenSSTX19} considers stratified sampling
design for both streaming and stored data, and improves upon the
Neyman allocation.  \revm{\cite{Haas:1992:SSP:130283.130335} estimates
  the size of a query result by partitioning the query result and
  estimating the sum of the partition sizes.  In our setting, each
  partition would be associated with one object, which contributes
  either $0$ or $1$ to the sum.  \cite{Haas:1992:SSP:130283.130335}
  focuses on deriving a sequential sampling procedure, and considers
  both uniform random sampling and stratified sampling.  However,
  unlike our work, it does not consider how to design stratification
  in a way to maximize sampling efficiency.}  Many other sampling
papers are concerned with aggregates such as \sql{SUM}, which are more
susceptible to sample biases than just counting queries.
\cite{DBLP:conf/icde/JoshiJ08} studies robust stratified sampling for
low-selectivity aggregate queries, and uses a pilot sampling phase to
estimate variance as we do for \SSN\ and \LSS.  A combination of
outlier-indexing with weighted sampling has been used in
\cite{DBLP:conf/icde/ChaudhuriDMN01} to approximate aggregate results,
and in~\cite{Babcock:2003:DSS:872757.872822}, where differently biased
subsamples can be dynamically selected to answer a query.
\cite{DBLP:journals/vldb/JoshiJ09} estimates the results of aggregates
over SQL queries with subqueries involving (\sql{NOT})
\sql{IN}/\sql{EXISTS}; notably, it proposes a low-variance estimator
by learning a model from data using Bayesian statistical techniques.
Compared with~\cite{DBLP:journals/vldb/JoshiJ09}, our approach is
simpler, uses off-the-shelf methods, and relies much less on the
quality of the learned models.  With the exception
of~\cite{DBLP:journals/vldb/JoshiJ09}, none of the work above applies
any machine learning to help with estimation or to inform the sampling
design.

\revb{Sampling has also been used for answering queries from dirty
  data with data cleaning~\cite{DBLP:conf/sigmod/WangKFGKM14}.  In
  Section~\ref{sec:prelim:correct} we have discussed this connection
  and introduced the method \QLSC\ inspired by
  \emph{SampleClean}~\cite{DBLP:conf/sigmod/WangKFGKM14}.  Experiments
  in Section~\ref{sec:expts} show that our learn-to-sample approach is
  more effective and less dependent on classifier quality.}

\paragraph{Use of Machine Learning}
\revm{There has been a flurry of recent research on the use of machine
  learning in database systems.  One related line of work is the use
  of machine learning for selectivity estimation,
  e.g.,~\cite{DBLP:conf/cidr/KipfKRLBK19,
    DBLP:conf/dasfaa/HalfordSM19}, which can be seen as approximate
  counting queries.  This line of work typically precomputes and
  maintains data summaries to support query optimization, or more
  ambitious optimizations across all components of a database system,
  e.g.,~\emph{SageDB}~\cite{DBLP:conf/cidr/KraskaABCKLMMN19}.  Since
  their goal is to use estimates for optimization instead of answering
  counting queries per se, their estimates typically do not come with
  any guarantees.  In contrast, we strive to provide statistical
  guarantees on our approximate answers.}

Finally, two recent papers are very similar to our approach in spirit.
\revb{To reduce the cost of evaluating expensive UDFs that arise
  frequently in machine learning pipelines, \emph{Probabilistic
    Predicates (PP)}~\cite{lu_accelerating_2018} use learned
  classifiers to pre-filter data before processing them further.
  Given a set of such classifiers and an accuracy requirement (minimum
  fraction of positives to retain), a query optimizer devises a plan
  that uses appropriate classifiers with optimal score cutoffs to
  pre-filter the data: any object scored lower than the cutoff by the
  classifier is dropped.  A key difference between this work and ours
  is the problem definition: they target \emph{reporting} queries
  while we target \emph{counting} queries.  This difference leads to
  our different use of the classifier scores; applying PP to our
  setting would result in poor estimates.  Furthermore, PP gives no
  statistical guarantees on the actual recall, and its performance is
  far more susceptible to bad classifiers because of its heavy
  reliance on classifier scores.}  Earlier work by Joglekar et
al.~\cite{joglekar_exploiting_2015} similarly tackles queries
involving selections with expensive UDFs.  By identifying attributes
whose values are correlated with UDF results, and grouping objects by
the values of such attributes, they judiciously choose the appropriate
actions to take for each group of objects (e.g., accept all, return
all, or sample some).  Like our approach, the use of sampling enables
probabilistic guarantees, but the key difference remains that they
target reporting instead of counting queries.

\section{Conclusion and Future Work}
\label{sec:conclude}

In this paper, we have developed new techniques to estimate the
results of counting queries with complex filters.  Our techniques are
based on a simple yet powerful idea: replace an expensive filter with
a cheap classifier that approximates the filter.  This cheap
classifier can then be used in a number of different ways with
different trade-offs.  \revm{A key challenge is that too much reliance
  on the classifier makes result quality highly susceptible to bad
  classifiers.  However, one novel technique we proposed,
  \emph{learned stratified sampling}, delivers consistently good
  estimates compared with other alternatives.  It is very resilient
  against bad classifiers, thanks to how it combines machine learning
  and sampling---the learned classifier is used in a limited but
  helpful way to design a stratified sampling scheme which in turn
  produces the estimate.  This resiliency makes the technique easy to
  apply in practice, because we are much less concerned with training
  a perfect model: a good model will make sampling more efficient, but
  even if the model is poor and/or the filter is fundamentally hard to
  approximate, the technique still delivers unbiased estimates with
  statistical guarantees comparable to random sampling.  There is an
  abundance of future work.  In particular, learned stratified
  sampling is quite conservative by design---to ensure independence,
  it avoids using the samples it acquired in the learning phase when
  computing the final estimate.  However, there may be ways in which
  such samples can be safely used.} Second, some of the queries we
considered in this paper (such as skyband sizes and neighbor counts)
have highly specialized solutions.  Although our goal is to develop
general solutions that can work for far more complex queries, it will
still be interesting to carry out a direct comparison with the
specialized solutions for these specific queries.  \revb{Finally, a
  promising direction is to extend and fully evaluate our approach in
  an \emph{online
    aggregation}~\cite{Hellerstein:1997:OA:253260.253291} setting.}

\balance
\bibliographystyle{abbrv}
\bibliography{bibliography}

\newpage

\begin{appendix}
\section{Details on \StraDirSol}
\label{apndxSec1}
We have
\begin{align*}
f(N_1, N_3)&=\frac{1}{n}[N_1s_1+N_3s_3+(N-N_1-N_3)s_2]^2\\
           &-[N_1s_1^2+N_3s_3^2+(N-N_1-N_3)s_2^2]\\
           &=\frac{(s_1-s_2)^2}{n}N_1^2 + \frac{(s_3-s_2)^2}{n}N_3^2+\\
           &\quad \frac{2(s_1-s_2)(s_3-s_2)}{n}N_1N_3\\
           &+[\frac{2(s_1-s_2)Ns_2}{n}-(s_1^2-s_2^2)]N_1\\
           &+[\frac{2(s_3-s_2)Ns_2}{n}-(s_3^2-s_2^2)]N_3+\frac{N^2s_2^2}{n}-Ns_2^2.
\end{align*}
In addition, recall the constraints on $N_1, N_3$ that we have and define a polygon $R$ on the plane with at most $5$ edges.


Let $f(N_1,N_3)=a_1N_1^2+a_2N_3^2+a_3N_1N_3+a_4N_1+a_5N_3+a_6$, where $a_1=\frac{(s_1-s_2)^2}{n}$,
$a_2=\frac{(s_3-s_2)^2}{n}$, $a_3=\frac{2(s_1-s_2)(s_3-s_2)}{n}$, $a_4=\frac{2(s_1-s_2)Ns_2}{n}-(s_1^2-s_2^2)$,
$a_5=\frac{2(s_3-s_2)Ns_2}{n}-(s_3^2-s_2^2)$, and $a_6=\frac{N^2s_2^2}{n}-Ns_2^2$.

Our goal is to minimize $f$ inside the polygon $R$ defined by the constraints on $N_1, N_3$.

Let $T$ be the list of the possible points that minimize $f$. We set $T=\emptyset$.

In order to minimize the quadratic function we find the critical points by computing the partial derivatives and set them to $0$.
$\frac{\theta f}{\theta N_1}=0\Leftrightarrow 2a_1N_1+a_3N_3+a_4=0$, and $\frac{\theta f}{\theta N_3}=0\Leftrightarrow 2a_2N_3+a_3N_1+a_5=0$.
We solve this linear system of two unknowns. We consider three cases. If the linear system has a unique solution $\bar{N}_1, \bar{N}_2$ then we check if
the point $(\bar{N}_1, \bar{N}_2)$ lies in $R$. If this is the case, we store
the point $(\bar{N}_1, \bar{N}_2)$ in $T$, otherwise we do not add any pair in $T$.
If the linear system has infinite solutions then it holds that in any solution $N_1', N_3'$ of the system,
$2a_1N_1'+a_3N_3'+a_4=0$. This function defines a line on the $2$-dimensional space of $N_1, N_2$. We check if
the line $2a_1N_1+a_3N_3+a_4=0$ intersects the $R$. If this is the case then let $q=(q_1,q_2)$ be one intersection point.
We add the point $(q_1,q_2)$ in $T$. Notice that $f$ takes the same value for all points on the line $2a_1N_1+a_3N_3+a_4=0$, hence it is sufficient
to store only one point. If the line does not intersect $R$ then we do not add any pair in the list. Finally, if the linear system has no solution
then we do not add any pair in the list.

So far we only searched for the critical points of the function $f$.
In case that those critical points do not lie in $R$ or if those points are saddle points or global maxima, the function $f$ is minimized
over the boundary of $R$. We continue our algorithm assuming the minimization of $f$ on the boundary of $R$.
Then, for each side of $R$ we do the following. We only describe it for the side where
$N_1=\max\{\Nmin, \imath_i\}$ and without loss of generality assume that $N_1=\imath_i$.
The rest of the sides of $R$ can be processed wit the same way.
Since, $N_1=\imath_i$, we have to minimize the function
$f(\imath_i, N_3)=a_2N_3^2+(a_3\imath_i+a_5)N_3+a_4\imath_i+a_1\imath_i^2+a_6$, which is a quadratic function with one variable. We can easily check
the minimum value of the function $f(\imath_i,N_3)$, by computing the derivative. Let $(\imath_i, \hat{N}_3)$ be the pair that minimizes $f(\imath_i, N_3)$. We add, $(\imath_i, \hat{N}_3)$ in $T$.

After computing $T$, we check all points in $T$ to find the optimum: Let $(x_1, x_3)$ be a point in $T$.
We evaluate the function $f$ on a point $(x_1', x_3')\in R$ which is the closest point to $(x_1,x_3)$ with integer coordinates.
In the end, we keep the integer values $(\bar{N}_1, \bar{N}_3)$ with the minimum $f(\bar{N}_1, \bar{N}_3)$ over all pairs in $T$.
We also have $\bar{N}_2=N-\bar{N}_1-\bar{N}_3$.

We repeat the above procedure for each pair of sampled points and in the end we return the boundaries that give the overall minimum variance.
You can see the pseudocode of our algorithm in Algorithm \ref{alg1}.

\begin{algorithm}
\caption{Pseudocode}\label{alg1}
\begin{algorithmic}[1]
\Procedure{Pseudocode}{}
   \State Sort $M$
   \State Compute the array $\SamplesPS$
   \State $(y_1^*, y_2^*)=\emptyset, r^*=+\infty$
   \For{each pair $(i, j)$ with $\mmin \le i < i+\mmin < j \le m-\mmin+1$}
   \State $s_1^2\textstyle= {\SamplesPS(i) \over i-1} \left( 1 -
          {\SamplesPS(i) \over i} \right)$
   \State $s_2^2=\textstyle= {\SamplesPS(j-1)-\SamplesPS(i) \over j-i-2}
          \left( 1 - {\SamplesPS(j-1)-\SamplesPS(i) \over j-i-1}
          \right)$
   \State $s_3^2=\textstyle= {\SamplesPS(m)-\SamplesPS(j-1) \over m-j}
          \left( 1 - {\SamplesPS(m)-\SamplesPS(j-1) \over m-j+1}
          \right)$
   \State Define polygon $R$ based on the constraints:
   \State \quad $\max\{\Nmin, \imath_i\} \le N_1 \leq \imath_{i+1}-1$,
   \State \quad $\max\{\Nmin, N-\imath_j+1\} \le N_3 \leq N-\imath_{j-1}$,
   \State \quad $N_1+N_3 \le N-\Nmin$
   \State $a_1=\frac{(s_1-s_2)^2}{n}$
   \State $a_2=\frac{(s_3-s_2)^2}{n}$
   \State $a_3=\frac{2(s_1-s_2)(s_3-s_2)}{n}$
   \State $a_4=\frac{2(s_1-s_2)Ns_2}{n}-(s_1^2-s_2^2)$
   \State $a_5=\frac{2(s_3-s_2)Ns_2}{n}-(s_3^2-s_2^2)$
   \State $a_6=\frac{N^2s_2^2}{n}-Ns_2^2$
   \State Define $f(N_1,N_3)=a_1N_1^2+a_2N_3^2+a_3N_1N_3+a_4N_1+a_5N_3+a_6$
   \State Let $T$ be the set of critical points of $f$ in $R$ along with the candidate solutions over each side of $R$
   \State $(y_1, y_3)=\emptyset$, $r=+\infty$
   \For{$(x_1, x_3)\in T$}
     \State Let $(x_1', x_3')$ be the closest point from $(x_1, x_3)$ with $x_1', x_3'\in \mathbb{Z}$, and $(x_1', x_3')\in R$
     \If{$f(x_1',x_3')<r$}
        \State $y_1=x_1', y_3=x_3'$
        \State $r=f(x_1',x_3')$
     \EndIf
   \EndFor
   \If{$f(y_1,y_3)<r^*$}
    \State $y_1^*=y_1, y_3^*=y_3$
    \State $r^*=f(y_1^*, y_3^*)$
   \EndIf
   \EndFor
   \State $\bar{N}_1=y_1^*$, $\bar{N}_3=y_3^*$, $\bar{N}_2=N-\bar{N}_1-\bar{N}_3$
   \State \Return $(\bar{N}_1, \bar{N}_2, \bar{N}_3)$
\EndProcedure
\end{algorithmic}
\end{algorithm}

Assume that $(x_1,x_3)$ is the pair that minimizes the function $f$ and $(N_1, N_3)$ is the closest integer coordinates point.
Furthermore, let $N_1^*, N_3^*$ be the optimum integer values that minimize the function $f$.
Let $v^*$ denote the minimum value of estimated variance defined in~\eqref{eq:stratified-obj:neyman}
  achievable using $n$ samples under stratified sampling with $H=3$
  strata where each stratum contains at least \Nmin\ objects.

\begin{lemma}
\label{lem:MainLem}
Assuming that $\Nmin > n$, \StraDirSol\ algorithm returns the boundaries of three strata with variance $v$, such that $v\leq (1+\frac{2}{\Nmin}+\frac{2}{\Nmin-n}+\frac{4}{\Nmin(\Nmin-n)})v^*$.
\end{lemma}
\begin{proof}
First, we observe that $|x_1-N_1|, |x_3-N_3|\leq 1$ (notice that the intersection of the line $N_1+N_3 \le N-\Nmin$ with the rectangle defined by
$\max\{\Nmin, \imath_i\} \le N_1 \leq \imath_{i+1}-1$, $\max\{\Nmin, N-\imath_j+1\} \le N_3 \leq N-\imath_{j-1}$ have integer coordinates so we can always find integer coordinates $N_i$ in $R$ within distance $1$).
Let also $x_2=N-x_1-x_3$, $N_2=N-N_1-N_3$ and $N_2^*=N-N_1^*-N_2^*$.
In the worst case if $N_1=x_1-1$ and $N_3=x_3-1$ then $N_2=x_2+2$.
For simplicity we analyze the function $g(z_1, z_2, z_3)=\frac{1}{n}(z_1s_1+z_2s_2+z_3s_3)^2-(z_1s_1^2+z_2s_2^2+z_3s_3^2)$, which is equivalent with the $f$ function. In particular, notice that $f(x_1,x_3)=g(x_1,x_2,x_3)$ and $f(N_1,N_3)=g(N_1, N_2, N_3)$.
We have $g(x_1, x_2, x_3)=f(x_1,x_3)\leq f(N_1^*, N_3^*)=g(N_1^*, N_2^*, N_3^*)$.
We compare $g(x_1, x_2, x_3)$ with $g(N_1, N_2, N_3)$.
We can re-write the function $g$ as
\begin{align*}
g(N_1,N_2,N_3)&=\sum_{i=1}^3 N_is_i^2(\frac{N_i}{n}-1)\\
              &+ 2\frac{N_1N_2s_1s_2+N_1N_3s_1s_3+N_2N_3s_2s_3}{n}
\end{align*}
We compare $g(x_1, x_2, x_3)$ with $g(N_1, N_2, N_3)$ term by term.


We have,
\begin{align*}
N_2s_2^2(\frac{N_2}{n}-1)&\leq (x_2+2)s_2^2(\frac{x_2+2}{n}-1)\\
                         &\hspace*{-3em}=x_2s_2^2(\frac{x_2}{n}-1)(1+\frac{2}{x_2}+\frac{2}{x_2-n}+\frac{4}{x_2(x_2-n)})\\
                         &\hspace*{-3em}\leq x_2s_2^2(\frac{x_2}{n}-1)(1+\frac{2}{\Nmin}+\frac{2}{\Nmin-n}+\frac{4}{\Nmin(\Nmin-n)}).
\end{align*}
With the same way, it is easy to observe that $N_1s_1^2(\frac{N_1}{n}-1)\leq x_1s_1^2(\frac{x_1}{n}-1)(1+\frac{2}{\Nmin}+\frac{2}{\Nmin-n}+\frac{4}{\Nmin(\Nmin-n)})$, and $N_3s_3^2(\frac{N_3}{n}-1)\leq x_3s_3^2(\frac{x_3}{n}-1)(1+\frac{2}{\Nmin}+\frac{2}{\Nmin-n}+\frac{4}{\Nmin(\Nmin-n)})$.

It remains to bound the terms $\frac{1}{n}N_1s_1N_2s_2, \frac{1}{n}N_1s_1N_3s_3$, and $\frac{1}{n}N_2s_2N_3s_3$.
We start by expanding the term $\frac{1}{n}N_1s_1N_2s_2$. We have,
\begin{align*}
\frac{1}{n}N_1s_1N_2s_2&\leq \frac{1}{n}(x_1+1)s_1(x_2+2)s_2\\
                       &=\frac{1}{n}x_1s_1x_2s_2(1+\frac{2}{x_2}+\frac{1}{x_1}+\frac{2}{x_1x_2})\\
                       &\leq \frac{1}{n}x_1s_1x_2s_2(1+\frac{2}{\Nmin}+\frac{2}{\Nmin-n}+\frac{4}{\Nmin(\Nmin-n)}).
\end{align*}
Similarly, it is easy to see that $\frac{1}{n}N_1s_1N_3s_3\leq \frac{1}{n}x_1s_1x_3s_3(1+\frac{2}{\Nmin}+\frac{2}{\Nmin-n}+\frac{4}{\Nmin(\Nmin-n)})$ and
$\frac{1}{n}N_1s_1N_3s_3\leq \frac{1}{n}x_1s_1x_3s_3(1+\frac{2}{\Nmin}+\frac{2}{\Nmin-n}+\frac{4}{\Nmin(\Nmin-n)})$.

We have that $g(x_1,x_2,x_3)\leq g(N_1^*, N_2^*, N_3^*)=v^*$, so we conclude that
$$g(N_1,N_2,N_3)\leq (1+\frac{2}{\Nmin}+\frac{2}{\Nmin-n}+\frac{4}{\Nmin(\Nmin-n)})g(N_1^*, N_2^*, N_3^*).$$
\end{proof}

That concludes the proof of Theorem~\ref{thm:DirSol}.

\section{Details on \StraLogBdr}
\label{apndxSec2}

\begin{lemma}
\label{lem:MainKLem2}
Assuming that $\Nmin> n$, \StraLogBdr\ algorithm returns the boundaries of $k$ strata with variance $v$, such that
$v\leq \max\{4, 2+2\max_{1\leq i\leq H}\frac{N_i^*}{N_i^*-n}\}v^*$, where $N_i^*$ is the size of the $i$-th stratum in the optimum solution.
\end{lemma}
\begin{proof}
We assume that the optimum allocation contains\\$N_1^*, \ldots, N_H^*$ points in stratum $1,\ldots, H$, respectively.
Let $S'$ be the set that contains the last sampled object in each stratum defined by $N_1^*, \ldots, N_H^*$.
Since the \StraLogBdr\ algorithm considers all possible ways to partition the sampled points it will definitely consider the partitioning where the sampled objects in $S'$ are the last sampled points in each stratum.
Let $t_1, \ldots, t_{H-1}$ be the boundaries of the optimum strata.
Let $t_i'$ be the leftmost point in $B_i$ which is at the right side of $t_i$.
Let $N_1, \ldots, N_H$ be the sizes of the strata we get using the points $t_i'$ as the boundary points.
We observe that $N_i\leq 2N_i^*$ for all $i$. Indeed, consider the size $N_i^*$.
Notice that in order to get $N_i$'s we always move the left boundaries of the strata to the right, so that makes the strata smaller.
Now we consider the right boundaries of the strata.
From the way that we constructed $B_i$ there always be a point $t_i'\in B_i$ such that
the number of points between $o_i$ and $t_i'$ is at most twice the number of points between $o_i$ and $t_i$.
In addition by using the boundary points $t_i'$ instead of the optimum $t_i$, notice that we do not change the
estimator $s_i$ for stratum $i$ for each $i\leq H$.

The rest of the proof is similar to the proof of Lemma \ref{lem:MainLem}.
Let $g(x_1, \ldots, x_H)$ be the function  as we defined in the proof of Lemma \ref{lem:MainLem}.
Let $N_1',\ldots, N_H'$ be the sizes that are found by our algorithm.
Since we return the boundaries with the smallest variance, it holds that $g(N_1',\ldots, N_H')\leq g(N_1, \ldots, N_H)$.
Finally, we compare $g(N_1,\ldots, N_H)$ with $g(N_1^*,\ldots, N_H^*)$, term by term as we did in the proof of Lemma \ref{lem:MainLem}.

For any pair $i, j$ we have $\frac{1}{n}N_is_iN_js_j\leq 4\frac{1}{n}N_i^*s_iN_j^*s_j$.
For any $i$, we also have,
$$N_is_i^2(\frac{N_i}{n}-1)\!\leq\!2N_i^*s_i^2(\frac{2N_i^*}{n}-1)\!\!=N_1^*s_1^2(\frac{N_1^*}{n}-1)(2+2\frac{N_1^*}{N_1^*-n}).$$

In any case we conclude that,
$$g(N_1',\ldots, N_H')\!\leq\!\max\{4, 2+2\max_{1\leq i\leq H}\frac{N_i^*}{N_i^*-n}\}g(N_1^*,\ldots, N_H^*).$$
\end{proof}

Notice that if $N_i^*\geq 2n$ then we have $v\leq 6v^*$.

As for running time, since the number of candidate boundary indices in
each of the $H-1$ sets is logarithmic in the number of objects between
two consecutive sampled object, the number of candidate
stratifications is $O(\log^{H-1} N)$.  Evaluating $V$ for each
candidate stratification takes $O(H)$ time, because each $s_h$ term
take constant time thanks to the prefix-sum index \SamplesPS\ (as in
\StraDirSol).  Overall, since there are $O(m^{H-1})$ possible
partitionings of $\SamplesI$ to consider, the total running time of
this algorithm, including precomputation time of $O(N \log m)$ (same
as \StraDirSol), is $O(N\log m + Hm^{H-1}\log^{H-1} N)$.

That concludes the proof of Theorem~\ref{thm:LogBdr}.

\section{Details on \StraDynPgm}
\label{apndxSec3}
We show the approximation factor we get from algorithm \StraDynPgm.
Let $N_1^*,\ldots, N_H^*$ be the cardinalities of the strata in the optimum solution with variance $v^*$.

\begin{lemma}
There exist strata with boundaries assuming the points in $B$ and sizes $\bar{N}_1,\ldots, \bar{N}_H$ such that $V(\bar{N}_1,\ldots, \bar{N}_H)\leq
\max\{4, 2+2\max_{1\leq i\leq H}\frac{N_i^*}{N_i^*-n}\}v^*$ and $2N^*_i\geq \bar{N}_i\geq N^*_i/2$ for each $i\leq H$.
\end{lemma}
\begin{proof}
Let $t_l$ be the right boundary of stratum $l$ in the optimum solution. Let $t_l'\in B$ be the leftmost object at the right of $t_l$.
We construct the strata with cardinalities $\bar{N}_1,\ldots,\bar{N}_k$ by considering the objects $t_l'$ for each $l$ as boundaries.
As we discussed in the slower algorithm for any $H$ we have
$V(\bar{N}_1,\ldots, \bar{N}_H)\leq
\max\{4, 2+2\max_{1\leq i\leq H}\frac{N_i^*}{N_i^*-n}\}v^*$ and $2N^*_i\geq \bar{N}_i$ for each $i\leq H$.
Assume that $p_r, p_{r+1}$ be the two consecutive points in $M$ such that $w(p_r)<w(t_l)<w(p_{r+1})$ (if $t_l\in M$ then the result follows).
If there are $y$ points in $P$ between $p_r$ and $t_l$ we have the guarantee that between $p_r$ and $t_l'$ there are at most $2y$ points.
Since $B$ contains all points in the order of powers of two from $o_{i+1}$ to $o_{i}$ it also follows that $\bar{N}_i\geq N^*_i/2$ for each $i\leq H$.
\end{proof}

For simplicity we consider that for each $i$, $N_i^* \geq 4n$. We can generalize the results even if $N_i^*$ is less than $4n$, however the analysis will be more tedious.
Hence, from the above lemma we have that $V(\bar{N}_1,\ldots, \bar{N}_H)\leq \frac{14}{3}v^*$.

Let $v'$ be the minimum variance of an allocation assuming the objects in $B$ when the size of the $l$-th stratum is $N_l\geq 2n$ for each $l\leq H$.
Notice that the solution $\bar{N}_1,\ldots, \bar{N}_H$ satisfies the inequalities $\bar{N}_l\geq 2n$ because $\bar{N}_l\geq N^*_i/2\geq 2n$, hence
the optimum allocation $N_1', \ldots, N_H'$ (and variance estimators $s_1',\ldots, s_H'$) using objects from $B$ with the additional constraint that $N_l'\geq 2n$ satisfies that $V(N_1',\ldots, N_H')\leq \frac{14}{3}v^*$.

Let $L=\sum_{i=1}^HN_i's_i'$ be the auxiliary sum of the optimum allocation in $B$ with the constraint $N_i'\geq 2n$.
We also assume that the $j$-th stratum with size $N_j'$ has its right boundary on $b_{i_j}\in B$.
We define $I=\{i_1, \ldots, i_H\}$.
Next, we prove that the dynamic programming algorithm returns a good approximation.

First, we assume that $\sum_{i=1}^HN_i's_i'\geq 1$.
Notice that $\sum_{i=1}^HN_i's_i'\leq \sum_{i=1}^HN_i'm\leq HNm$, so $HNm$ is an upper bound on any auxiliary sum.
Hence, in $T$ there should always be a value $t$ such that $L\leq t\leq 2L$.
Furthermore,
let $$O[i_j,j]=\frac{1}{n}(\sum_{l=1}^j N_l's_l')^2- \sum_{l=1}^{j} N_l'(s_l')^2.$$
\begin{lemma}
\label{apnd:lem1}
If $L\geq 1$, for each $i_j\in I$ we have that $A_t[i_j,j]\leq (8j-7)O[i_j,j]+4\frac{1}{n}\sum_{y=1}^j(y-1)N_y's_y'\sum_{l=y+1}^HN_l's_l'$.
\end{lemma}
\begin{proof}
We prove it by induction on the number of buckets. For $j=1$ we have from the definition that $A_t[i_1,1]$ has the optimum variance so
$A_t[i_1,1]= \frac{1}{n}(N_1's_1')^2-N_1'(s_1')^2$, and the inequality holds.
Assume that this is true for $A_t[i_{j-1},j-1]$. We will prove it for $A_t[i_j,j]$.
We have $$A_t[i_j,j]\!\leq\!\frac{1}{n}N_j'^2\!s_j'^2\!-\!N_j's_j'^2\!+\!A_t[i_{j-1}, j\!-\!1]\!+\!\frac{2}{n}N_j's_j'X_t[i_{j-1},j\!-\!1].$$
Since at each step we consider a stratum with $N_j\cdot s_j\leq t$, we have that
$X_t[i_{j-1},j-1]\leq (j-1)t\leq 2(j-1)\sum_{l=1}^HN_l's_l'$ so
\begin{align*}
A_t[i_j,j]&\leq \frac{1}{n}N_j'^2s_j'^2-N_j's_j'^2 + A_t[i_{j-1}, j-1]\\
          &\hspace*{3em}+ \frac{4(j-1)}{n}N_j's_j'\sum_{l=1}^HN_l's_l'.
\end{align*}
In addition we have,
$$N_j's_j'\sum_{l=1}^HN_l's_l'=N_j's_j'\sum_{l=1}^{j-1}N_l's_l' + (N_j')^2(s_j')^2+\\N_j's_j'\sum_{l=j+1}^HN_l's_l',$$
so
\begin{align*}
A_t[i_j,j]&\leq \frac{1}{n}N_j'^2s_j'^2-N_j's_j'^2 + A_t[i_{j-1}, j-1]\\
          &\hspace*{2em}+ \frac{4(j-1)}{n}(N_j's_j'\sum_{l=1}^{j-1}N_l's_l' + (N_j')^2(s_j')^2\\
          &\hspace*{12em}+ N_j's_j'\sum_{l=j+1}^HN_l's_l').
\end{align*}

We focus on the expression $\frac{1}{n}N_j'^2s_j'^2-N_j's_j'^2+\frac{4(j-1)}{n}(N_j')^2(s_j')^2$.
This can be written as $N_j'(s_j')^2(\frac{(4j-3)N_j'}{n}-1)$. Since $N_j'\geq 2n$ we have that $\frac{(4j-3)N_j'}{n}-1\leq (8j-7)(\frac{N_j'}{n}-1)$,
so
$$\frac{1}{n}N_j'^2s_j'^2-N_j's_j'^2+\frac{4(j-1)}{n}(N_j')^2(s_j')^2\leq (8j-7)(\frac{1}{n}N_j'^2s_j'^2-N_j's_j'^2).$$
Finally, $$2(j-1)\frac{1}{n}2N_j's_j'\sum_{l=1}^{j-1}N_l's_l'\leq (8j-7)\frac{1}{n}2N_j's_j'\sum_{l=1}^{j-1}N_l's_l',$$
and $$A_t[i_{j-1},j-1]\leq (8j-15)O[i_{j-1},j]+4\frac{1}{n}\sum_{y=1}^{j-1}(y-1)N_y's_y'\sum_{l=j}^HN_l's_l'$$ so we conclude that
$$A_t[i_j,j]\leq (8j-7)O[i_j,j]+4\frac{1}{n}\sum_{y=1}^j(y-1)N_y's_y'\sum_{l=y+1}^HN_l's_l'.$$
\end{proof}

From the lemma above, we have,
\begin{align*}
&A_t[i_H,H]\\
          &\hspace*{1em}\leq (8H-7)O[i_H,H]+\frac{1}{n}\sum_{y=1}^H2(y-1)(2N_y's_y'\sum_{l=y+1}^HN_l's_l')\\
          &\hspace*{1em}\leq (8H-7)O[i_H,H]+2(H-1)\frac{1}{n}\sum_{y=1}^H(2N_y's_y'\sum_{l=y+1}^HN_l's_l').
\end{align*}
Notice that
\begin{align*}
O[i_H,H]&=\frac{1}{n}(\sum_{a=1}^H N_a's_a')^2- \sum_{a=1}^{H} N_a'(s_a')^2\\
        &=\sum_{a=1}^HN_a'(s_a')^2(\frac{N_a'}{n}-1)+\frac{2}{n}\sum_{y=1}^H(N_y's_y'\sum_{l=y+1}^HN_l's_l'),
\end{align*}
where\\
$\sum_{a=1}^HN_a'(s_a')^2(\frac{N_a'}{n}-1)\geq 0$ and $\frac{2}{n}\sum_{y=1}^H(N_y's_y'\sum_{l=y+1}^HN_l's_l')\geq 0$.
Each term $\frac{2}{n}N_y's_y'N_l's_l'$, for $l$ such that $y+1\leq l$ exists in $O[i_H,H]$,
so we have $$A_t[i_H,H]\leq (10H-9)O[i_H,H]=(10H-9)V(N_1',\ldots, N_H').$$
Since $V(N_1',\ldots, N_H')\leq \frac{14}{3}v^*$, we conclude that
the solution we get in the dynamic programming algorithm has variance at most $\frac{14}{3}(10H-9)v^*$.

Finally, we assume that $\sum_{i=1}^HN_i's_i'< 1$. The proof is similar to the previous case.
Notice that in $T$ there should always be a value $t$ such that $L\leq t\leq L+\varepsilon$.
Similar to Lemma~\ref{apnd:lem1} we can show:
\begin{lemma}
If $L< 1$, for each $i_j\in I$ we have that $A_t[i_j,j]\leq (4j-3)O[i_j,j]+2\frac{1}{n}\sum_{y=1}^j(y-1)N_y's_y'\sum_{l=y+1}^HN_l's_l' + \frac{2\varepsilon}{n}\sum_{y=1}^j(y-1)N_y's_y'$.
\end{lemma}
\begin{proof}
We prove it by induction on the number of buckets. For $j=1$ we have from the definition that $A_t[i_1,1]$ has the optimum variance so
$A_t[i_1,1]= \frac{1}{n}(N_1's_1')^2-N_1'(s_1')^2$, and the inequality holds.
Assume that this is true for $A_t[i_{j-1},j-1]$. We will prove it for $A_t[i_j,j]$.
We have $$A_t[i_j,j]\!\leq\!\frac{1}{n}N_j'^2\!s_j'^2\!-\!N_j's_j'^2\!+\!A_t[i_{j-1}, j-1]\!+\!\frac{2}{n}N_j's_j'X_t[i_{j-1},j\!-\!1].$$
Since at each step we consider a stratum with $N_j\cdot s_j\leq t$, we have that
$X_t[i_{j-1},j-1]\leq (j-1)t\leq (j-1)(\sum_{l=1}^HN_l's_l'+\varepsilon)$ so
\begin{align*}
A_t[i_j,j]&\leq \frac{1}{n}N_j'^2s_j'^2-N_j's_j'^2 + A_t[i_{j-1}, j-1]\\
          &\hspace{2em}+ \frac{2(j-1)}{n}N_j's_j'\sum_{l=1}^HN_l's_l' + \frac{2\varepsilon(j-1)}{n}N_j's_j'.
\end{align*}
In addition we have,
$$N_j's_j'\sum_{l=1}^HN_l's_l'=N_j's_j'\sum_{l=1}^{j-1}N_l's_l' + (N_j')^2(s_j')^2+\\N_j's_j'\sum_{l=j+1}^HN_l's_l',$$
so
\begin{align*}
A_t[i_j,j]&\leq \frac{1}{n}N_j'^2s_j'^2-N_j's_j'^2 + A_t[i_{j-1}, j-1]\\
          &\hspace*{2em}+\frac{2(j-1)}{n}(N_j's_j'\sum_{l=1}^{j-1}N_l's_l' + (N_j')^2(s_j')^2\\
          &\hspace*{10em}+ N_j's_j'\sum_{l=j+1}^HN_l's_l')\\
          &\hspace*{2em}+ \frac{2\varepsilon(j-1)}{n}N_j's_j'.
\end{align*}

We focus on the expression $\frac{1}{n}N_j'^2s_j'^2-N_j's_j'^2+\frac{2(j-1)}{n}(N_j')^2(s_j')^2$.
This can be written as $N_j'(s_j')^2(\frac{(2j-1)N_j'}{n}-1)$. Since $N_j'\geq 2n$ we have that $\frac{(2j-1)N_j'}{n}-1\leq (4j-3)(\frac{N_j'}{n}-1)$,
so
$$\frac{1}{n}N_j'^2\!s_j'^2\!-\!N_j's_j'^2\!+\!\frac{2(j-1)}{n}(N_j')^2(s_j')^2\!\leq\!(4j-3)(\frac{1}{n}N_j'^2s_j'^2\!-\!N_j's_j'^2).$$
Finally, $$(j-1)\frac{1}{n}2N_j's_j'\sum_{l=1}^{j-1}N_l's_l'\leq (4j-3)\frac{1}{n}2N_j's_j'\sum_{l=1}^{j-1}N_l's_l',$$
and
\begin{align*}
A_t[i_{j-1},\!j\!-\!1\!]&\!\leq\!(4j-7)O[i_{j-1},j]\!+\!2\frac{1}{n}\!\sum_{y=1}^{j-1}(y\!-\!1\!)N_y's_y'\!\sum_{l=j}^H\!N_l's_l'\\
                &\hspace*{3em}+ \frac{2\varepsilon}{n}\sum_{y=1}^{j-1}(y-1)N_y's_y'
\end{align*}
so we conclude that
\begin{align*}
A_t[i_j,j]&\leq (4j-3)O[i_j,j]+2\frac{1}{n}\sum_{y=1}^j(y-1)N_y's_y'\sum_{l=y+1}^HN_l's_l'\\
          &\hspace*{3em}+\frac{2\varepsilon}{n}\sum_{y=1}^{j}(y-1)N_y's_y'.
\end{align*}
\end{proof}

From the lemma above, we have,
\begin{align*}
A_t[i_H,H]&\leq (4H-3)O[i_H,H]+\frac{1}{n}\sum_{y=1}^H2(y-1)(N_y's_y'\sum_{l=y+1}^HN_l's_l')\\
          &\hspace*{3em}+\frac{2\varepsilon}{n}\sum_{y=1}^{H}(y-1)N_y's_y'\\
          &\leq (4H-3)O[i_H,H]+(H-1)\frac{1}{n}\sum_{y=1}^H(2N_y's_y'\sum_{l=y+1}^HN_l's_l')\\
          &\hspace*{3em}+\frac{2\varepsilon}{n}\sum_{y=1}^{H}(y-1)N_y's_y'.
\end{align*}
Notice that $\frac{2\varepsilon}{n}\sum_{y=1}^{H}(y-1)N_y's_y'\leq 2\varepsilon$ since $H<n$, and $\sum_{y=1}^{H}N_y's_y'<1$.
In addition, notice that
\begin{align*}
O[i_H,H]&=\frac{1}{n}(\sum_{a=1}^H N_a's_a')^2- \sum_{a=1}^{H} N_a'(s_a')^2=\sum_{a=1}^HN_a'(s_a')^2(\frac{N_a'}{n}-1)\\
        &\hspace*{3em}+\frac{2}{n}\sum_{y=1}^H(N_y's_y'\sum_{l=y+1}^HN_l's_l'),
\end{align*}
where
$\sum_{a=1}^HN_a'(s_a')^2(\frac{N_a'}{n}-1)\geq 0$ and $\frac{2}{n}\sum_{y=1}^H(N_y's_y'\sum_{l=y+1}^HN_l's_l')\geq 0$.
Each term $\frac{2}{n}N_y's_y'N_l's_l'$, for $l$ such that $y+1\leq l$ exists in $O[i_H,H]$,
so we have $$A_t[i_H,H]\leq (5H-4)O[i_H,H]+\varepsilon=(5H-4)V(N_1',\ldots, N_H')+\varepsilon.$$
Since $V(N_1',\ldots, N_H')\leq \frac{14}{3}v^*$, we conclude that
the solution we get in the dynamic programming algorithm has variance at most $\frac{14}{3}(5H-4)v^*+\varepsilon$.

This concludes the proof of Theorem~\ref{thm:DynPgm}.

\section{Details on \PropDynPgm}
\label{apndxSec4}

\begin{lemma}
\PropDynPgm algorithm returns the boundaries of $k$ strata with variance $v$, such that
$v\leq 2v^*$
\end{lemma}
\begin{proof}
\PropDynPgm algorithm returns the optimum answer with respect to points in $B$, since the objective function is decomposable.
Let $N_1^*, \ldots, N_H^*$ be the sizes of the optimum strata.
Because of the selection of $B$ and Lemma \ref{lem:MainKLem2}, the dynamic programming algorithm always consider a solution where the size of each strata
is $N_h'\leq 2N_h^*$ for each $h\leq H$, and the estimated variance in each stratum is the same as in the optimum solution.
Hence, we can easily observe that $N_h's_h^2\leq 2N_h^*s_h^2$.
Let $v'$ be the variance of such a solution (with sizes $N_1',\ldots, N_H'$). Since $B$ returns the best solution we have that $v\leq v'\leq 2v^*$.
\end{proof}

That concludes the proof of Theorem~\ref{thm:PropDynPgm}. 
\end{appendix}

\end{document}